\documentclass[a4paper,fleqn,usenatbib]{mnras}

\usepackage{newtxtext,newtxmath}

\usepackage[T1]{fontenc}
\usepackage{ae,aecompl}

\usepackage{graphicx}	
\usepackage{amsmath}	
\usepackage{amssymb}	

\title[Sub-Keplerian rotation around L$_2$ Pup]{Radiation-pressure-driven sub-Keplerian rotation of the disc around the AGB star L$_2$ Pup. }

\author[T. J. Haworth et al. ]
{\parbox{\textwidth}{Thomas J. Haworth$^{1}$\thanks{E-mail: \texttt{t.haworth@imperial.ac.uk}}, Richard A. Booth$^{2}$, Ward Homan$^{3}$, Leen Decin$^{3}$, \\Cathie J. Clarke$^{2}$ and Subhanjoy Mohanty$^{1}$
}\vspace{0.4cm}\\
\parbox{\textwidth}{$^{1}$  Astrophysics Group, Imperial College London, Blackett Laboratory, Prince Consort Road, London SW7 2AZ, UK\\
$^{2}$ Institute of Astronomy, Madingley Rd, Cambridge, CB3 0HA, UK \\
$^{3}$ Institute of Astronomy, KU Leuven, Celestijnenlaan 200D B2401, 3001 Leuven, Belgium \\
}}

\date{Accepted XXX. Received YYY; in original form ZZZ}

\pubyear{2015}

\begin{document}
\label{firstpage}
\pagerange{\pageref{firstpage}--\pageref{lastpage}}
\maketitle

\begin{abstract}
We study the sub-Keplerian rotation and dust content of the circumstellar material {around} the {asymptotic giant branch (AGB)} star L$_2$ Puppis. We find that the thermal pressure gradient alone cannot explain the observed rotation profile. We find that there is a family of possible dust populations for which radiation pressure can drive the observed sub-Keplerian rotation. This set of solutions is further constrained by the {spectral energy distribution (SED)} of the system, and we find that a dust-to-gas mass ratio of $\sim10^{-3}$ and a {maximum grain size that decreases radially outwards} can satisfy both the rotation curve and SED.  These dust populations are dynamically tightly coupled to the gas azimuthally. However grains larger than $\sim0.5\mu$m are driven outward radially by radiation pressure at velocities $\sim5\,$km\,s$^{-1}$, which implies a dust replenishment rate of $\sim3\times10^{-9}$\,M$_\odot$\,yr$^{-1}$. This replenishment rate is consistent with observational estimates to within uncertainties. {Coupling between the radial motion of the dust and gas is weak and hence the gas does not share in this rapid outward motion.} Overall we conclude that radiation pressure is a capable and necessary mechanism to explain the observed rotation profile of L$_2$ Pup, and offers other additional constraints on the dust properties.

\end{abstract}




\section{Introduction}

{Low to intermediate-mass stars (stars with masses $< 8 M_\odot$ at solar metallicity)} ascend the {asymptotic giant branch} (AGB) on the Hertzsprung-Russell diagram (HRD) as they reach the end of their lives. {During the AGB phase, a combination of} surface pulsations, enabling the formation of dust, and radiation pressure on this dust is believed to drive a strong stellar wind, with typical velocities of $\sim 10$\,km/s \citep{2003agbs.conf.....H}. The mass lost in this wind ranges from $10^{-8}$ up to $10^{-4}$\,M$_\odot$\,yr$^{-1}$ \citep[e.g.][]{2010A&A...523A..18D} resulting in an important contribution to the gaseous and dusty enrichment of the interstellar medium (ISM). {A detailed understanding of the physical and chemical processes of such a wind (composition, mass loss rate, etc.) can hence provide better insight into the impact of AGB stars on global galactic chemical evolution (e.g. in terms of metallicity and dust-to-gas mass ratio).}

Recent high angular resolution observations of AGB circumstellar envelopes have shown that these winds harbour a wealth of structural complexities, ranging from small-scale clumps \citep{2016A&A...591A..70K} and arcs \citep{2015A&A...574A...5D} to large-scale spirals \citep{2012Natur.490..232M} and shells \citep{2015A&A...575A..91C}. The true origin of these morphologies is still a large point of debate, but it is generally believed that cylindrically-shaped morphologies, like spirals and equatorial density enhancements, materialise through wind-binary interactions \citep[e.g.][]{1993MNRAS.265..946T,1998ApJ...497..303M, 2012ApJ...759...59K,2016MNRAS.460.4182C}. The fraction of AGB circumstellar envelopes exhibiting such structures can be high, since the multiplicity frequency of the progenitors of AGB stars has been shown to be above 50 percent \citep{2010ApJS..190....1R,2013ARA&A..51..269D}. In addition, they form an important class of candidates that may explain the first stages in the morphological evolution from spherical stellar systems to the predominantly bipolar post-AGB stars and planetary nebulae.

In order to better understand the impact of binary effects on wind shaping, and by extension on the global (thermo)dynamical and chemical properties of AGB circumstellar envelopes, better theoretical and observational constraints are required. An ideal candidate for such in-depth exploration of the complete anatomy of an equatorial density enhancement is the recently discovered differentially rotating gas and dust {disc} around the AGB star L$_{\rm 2}$ Puppis \citep{2014A&A...564A..88K,2016A&A...596A..92K}.

L$_{\rm 2}$ Pup is a semi-regular pulsating variable with a period of $P=141$\,days \citep{1985IBVS.2681....1K,2005MNRAS.361.1375B}, an effective temperature of $T_\mathrm{eff} \approx 3500$\,K, and a radial velocity relative to the Local Standard of Rest (lsr) of $v_{\rm lsr} = 33.3$\,km\,s$^{-1}$. It is the second nearest asymptotic giant branch (AGB) star, located at a distance of only 64\,pc {\citep[$\pi = 15.61 \pm 0.99$\,mas; ][]{2007A&A...474..653V}}.

\cite{2016A&A...596A..92K} used ALMA to accurately probe the kinematics of the gas contained within this disc, detecting both Keplerian and sub-Keplerian motion in the equatorial plane of the disc. The Keplerian motion of the inner {disc} has permitted the very accurate determination of the mass of the central star, being 0.659$\pm$0.052 Solar masses. The {azimuthal (rotational) velocity} transitions from Keplerian to sub-Keplerian at the dust detection radius, strongly suggesting that the {properties of dust (dust-to-gas mass ratio, size distribution, etc.) in the} {disc} influences the local dynamics. \cite{2017A&A...601A...5H} have modelled the molecular $^{\rm 12}$CO and $^{\rm 13}$CO emission with 3D radiative transfer, inferring the gaseous density, temperature and velocity structure of the disc. A {likely} companion has also been detected, located at the inner rim of the gas disc, suggesting it plays a role in the formation of the equatorial structure.

In this paper we aim to further extend our understanding of the structure of discs surrounding evolved stars by investigating whether the sub-Keplerian motion observed in the {disc} {around} L$_{\rm 2}$ Pup could be induced by radiation pressure on the dust. {Indeed there is the expectation that radiation pressure influences dust-gas dynamics from prior work on massive stellar discs \citep[e.g.][]{2001ApJ...557..990T} and the circumstellar medium of post-AGB stars \citep[e.g.][]{2003A&A...397..595D}.} {In particular \cite{2003A&A...397..595D} found that radiation pressure could liberate small ($<10\mu$m) grains from high $z$ in the disc around the post-AGB star/binary companion HR\,4049, as well as affecting the radial drift of grains. However they did not consider the dynamical effect of this on the gas. L$_{\rm 2}$ Pup also differs from HR\,4049 in that it has a much lower disc mass, of order $10^{-4}-10^{-3}$\,M$_{\odot}$ compared to  0.3\,M$_{\odot}$ \citep{2017A&A...601A...5H}. The lower optical depth of L$_{\rm 2}$ Pup means that the effect of radiation on dust could be much more pervasive.  }

{In addition to understanding the role of radiation pressure in driving sub-Keplerian rotation of the disc around  L$_{\rm 2}$ Pup, we also aim} to determine what dust grain species and grain size (distributions) populate the disc, and by extension the stellar outflow. 
Ultimately we aim to contribute to a deeper understanding of the formation and stability of AGB circumstellar discs. Furthermore, better understanding the inner circumstellar envelopes of AGB stars will improve our understanding of the mechanisms that drive the AGB wind itself, and the subsequent evolutionary steps towards the AGB progeny: the post-AGB stars and planetary nebulae (PN), whose global morphology deviates significantly from the spherically symmetric AGB predecessors.

\section{Radiation pressure induced sub-Keplerian rotation}
\subsection{Basic concept}
\label{sec:concept}
We propose that the sub-Keplerian rotation identified in the {disc} {around} L$_2$ Pup might be explained by radiation pressure. Assuming that dust and gas are well dynamically coupled, and that the dust can exert a dynamical back-reaction on the gas through momentum conservation, we can trivially extend the balancing of centrifugal force and gravity that is Keplerian rotation to include radiation pressure. This yields a steady state azimuthal  rotation profile as a function of radial distance $R$ of
\begin{equation}
	v_{\phi} = \sqrt{\frac{GM_*}{R} - \frac{f_{\textrm{rad}}R}{\rho}}
	\label{vphi}
\end{equation}
where $f_{\textrm{rad}}$ is the radiation pressure force per unit volume, $M_*$ is the stellar mass, $\rho$ is the local volume density and $G$ the gravitational constant. The assumed dynamic coupling is valid for grains with Stokes numbers (the ratio of grain stopping time to dynamical time-scale) much less than unity. We address this assumption in section \ref{dynmodel}. 
A thermal pressure gradient $dP/dR$ will also support against Keplerian rotation. Accounting for this extends equation \ref{vphi} to
\begin{equation}
	v_{\phi} = \sqrt{\frac{GM_*}{R} - \frac{f_{\textrm{rad}}R}{\rho} + \frac{R}{\rho}\frac{dP}{dR}}.
	\label{vphi_wpgrad}
\end{equation}
where the mid-plane radial pressure gradient will be negative.

\begin{figure}
	\includegraphics[width=9.5cm]{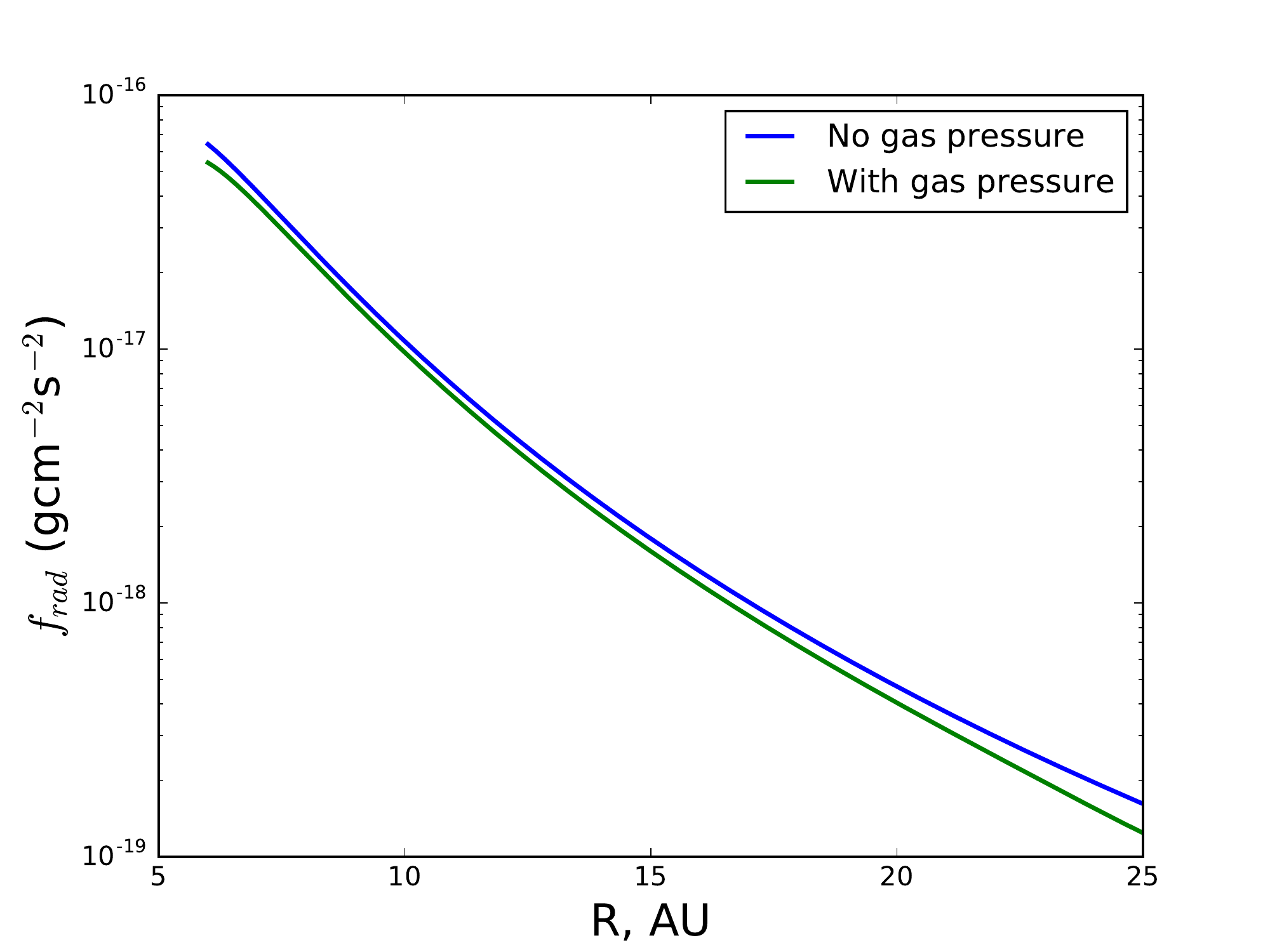}
	\caption{The radiation pressure force per unit volume required to give the deviation from Keplerian velocity observed towards L$_2$ Pup, both with and without accounting for the radial pressure gradient. Note that this is for the {disc} mid-plane.}
	\label{freq}
\end{figure}

\cite{2016A&A...596A..92K} summarised the observed azimuthal velocity profile of L$_2$ Pup  as
\begin{equation}
	v_{\phi} = 40.7\left(\frac{R}{\textrm{AU}}\right)^{-0.853}\,\textrm{km\,s}^{-1}.
	\label{vphiobs}
\end{equation}
The models of \cite{2017A&A...601A...5H} imply that the mid-plane density distribution is 
\begin{equation}
	\rho_{\textrm{mid}} = 9.3\times10^{-13}\left(\frac{R}{R_c}\right)^{-3.1}\,\textrm{g\,cm}^{-3}
	\label{rhomid}
\end{equation}
where $R_c=2$\,AU.
We can combine equations \ref{vphi_wpgrad}, \ref{vphiobs} and \ref{rhomid} to solve for the $f_{\textrm{rad}}$ required to give the observed velocity distribution, in the mid-plane,
\begin{equation}
	f_{\textrm{rad}} = \frac{\rho}{R}\left(v_{\textrm{kep}}^2 - v_{\phi}^2\right) +\frac{dP}{dR}
\end{equation}
where $v_{\textrm{kep}}$ is regular Keplerian rotation and $\rho$ the local density. This required mid-plane $f_{\textrm{rad}}$ profile for the {disc} {around} L$_2$ Pup, both with and without the mid-plane radial pressure \citep[using the thermal description of the {disc} computed by][which we present in section \ref{discConstruction}, equation \ref{homanTemp}]{2017A&A...601A...5H} is shown in Figure \ref{freq}.

Note that this relation suggests the existence of a solution for which $v_\phi=0$. This is equivalent to the critical point at which the radiation pressure balances the inward gravitational pull ($f_{\textrm{rad}}\approx \rho v_{\textrm{kep}}^2/R$). Radiation pressures exceeding this value would not permit stable orbits. The resulting particle trajectories are then likely to be purely radially outward.

Given that equation \ref{vphi} tells us the steady state azimuthal velocity as a function of radiation pressure, we can use a radiative transfer code to estimate the radiation pressure and hence the steady state rotational profile for different configurations, without having to perform a full radiation hydrodynamic simulation. This is our focus for the rest of this paper.

\section{Numerical method}
\label{nummethod}
We now summarise our radiative transfer models used to probe the dust distribution and sub-Keplerian rotation of the {disc} {around} L$_2$ Pup. 

\subsection{Model parameters}
\label{modelparams}
We use the Monte Carlo radiation transport code \textsc{torus} for the calculations in this paper \citep[e.g.][]{2000MNRAS.315..722H, 2012MNRAS.420..562H, 2015MNRAS.448.3156H}. \textsc{torus} computes the radiation pressure force using the algorithm presented by \cite{2015MNRAS.448.3156H}. This method treats polychromatic radiation and anisotropic scattering in the free streaming and optically thick limits. In short the photon source (stellar) luminosity is broken into discrete, constant energy, packets of photons which are propagated through the computational domain on a random walk -- much like photons propagating through a medium in reality. As each packet of energy $\epsilon_i$ traverses a path length $l$ through a cell it contributes to the radiation pressure force in that cell. Once all packets are propagated, the total radiation pressure force per unit volume in cell $j$ is
\begin{equation}
	f_{\textrm{rad}, j} = \frac{1}{c}\int\kappa_\nu\rho F_{\nu}d\nu = \frac{1}{c}\frac{1}{\Delta t}\frac{1}{V_j}\sum \epsilon_i \kappa_\nu \rho l \hat{u}
	\label{radfequn}
\end{equation}
where $\kappa_{\nu}$, $\rho$, $V$ and $\hat{u}$ are the cell specific dust opacity, density, cell volume and the radiation pressure force unit vector respectively. In this paper we do not perform full radiation hydrodynamic simulations, which are computationally expensive (\textsc{torus} also currently assumes dynamically coupled dust and gas). Rather we compute the radiation pressure force and feed it into the analytic framework discussed above. 

Monte Carlo radiative transfer with \textsc{torus} is also used to compute synthetic {spectral energy distributions} (SEDs) from our models. Note that for SEDs we directly compute the dust radiative equilibrium temperature using an approach based on \cite{1999A&A...344..282L}. That is, we don't use the parametric temperature for the {disc} derived by \cite{2017A&A...601A...5H}, but calculate it explicitly. 

\textsc{torus} permits the use of multiple dust types across different spatial regions in a given simulation, where a particular dust type has a dust-to-gas mass ratio ($\delta$) minimum/maximum grain size ($a_{\textrm{min}}$, $a_{\textrm{max}}$), a power law distribution ($q$) and a composition. For the models in this paper we use 10 different dust types that apply over discrete radial ranges (e.g. the first spans from 6--7\,AU). The dust parameters are not allowed to vary vertically at this stage, but given that \cite{2017A&A...601A...5H} inferred a turbulent velocity of $\sim1$\,km/s, which would result in vertical mixing of the contents of the {disc} on times of the order of a few tens of years, the assumption of vertically constant dust properties is prudent. 

Unless otherwise stated, the grain compositions that we consider in our models are \cite{2003ApJ...598.1017D} silicates. In section \ref{grainComposition} we also consider the iron poor Mg(0.95) Fe(0.05) SiO(3) and iron rich Mg Fe SiO(4) magnesium-iron silicates, with data from the Jena \textsc{doccd} database\footnote{\url{http://www.astro.uni-jena.de/Laboratory/OCDB/amsilicates.html}} \citep{1994A&A...292..641J, 1995A&A...300..503D}. The optical constants of these grain types are used to compute a Mie scattering phase matrix.

We assume an \cite{1977ApJ...217..425M} size distribution, $\textrm{d}n/\textrm{d}a \propto a^{-q}$, of grains between the minimum and maximum grain size in each radial bin. The radial variation of these dust parameters is to be determined, such that the azimuthal velocity is consistent with that observed. Note that our calculations are insensitive to the gas composition because grains will dominate the opacity in continuum radiative transfer.

In this paper we locate solutions to the rotation profile manually. That is we make an initial guess of the dust properties, calculate the rotation profile and then modify, e.g. $a_{\textrm{max}}$ in each bin to drive the solution  towards the observed rotation profile. Once one solution is found, say for a fixed dust-to-gas ratio, generating others for small deviations in the dust-to-gas ratio is done relatively quickly given that small perturbations to the first solution are required. 

\subsection{Stellar model}
\label{sec:source}
We model the stellar spectrum of L$_2$ Pup using the models of \cite{2004astro.ph..5087C}. However, these only extend into the far-infrared out to $\sim160$\,$\mu$m. Beyond this wavelength the emission is very similar to a blackbody spectrum, which we therefore adopt beyond the bounds of the more sophisticated spectral models. We assume an effective temperature of 3500\,K, a luminosity of 2000L$_{\odot}$, a radius of 121\,R$_{\odot}$ and a mass of 0.659\,M$_{\odot}$ \citep{2016A&A...596A..92K}. Note that we do not account for radiation from the possible secondary under the assumption that the AGB star dominates. 

\subsection{Disc construction}
\label{discConstruction}
We base our {disc} on the best-fit models of \cite{2017A&A...601A...5H}. The gas density is set by
\begin{equation}
\rho = \rho_0\left(\frac{r}{R_c}\right)^{-3.1}\exp\left(-\frac{z^2}{2H^2}\right)
    \label{rhoEqun}
\end{equation}
where $r=\sqrt{x^2+y^2}$, $R_c=2$\,AU,  $\rho_0=9.3\times10^{-13}$\,g\,cm$^{-3}$ and
\begin{equation}
    H = H_c\left(\frac{r}{Rc}\right)^{0.2}
\end{equation}
where $H_c=1.5$\,AU. We impose a {disc} outer radius of 26\,AU {\citep[approximately the observed extent: ][]{2016A&A...596A..92K, 2017A&A...601A...5H}} beyond which we set the density to a negligbly low value. At this stage we only permit the dust properties to vary radially, not vertically. We always impose a dust-free inner 6\,AU, as expected from observations. When computing the radiation pressure force we assume the background thermal structure concluded by \cite{2017A&A...601A...5H}, which is 
\begin{equation}
    T = \left(T_z-T_p\right)\exp\left(-\frac{r^2}{2w_1^2}\right) - \left(T_p/\pi\right)\tan^{-1}\left(\frac{r-D}{w2} - \frac{\pi}{2}\right)
    \label{homanTemp}
\end{equation}
where $T_z=2500$\,K, $T_p=500$\,K, $w_1=1.8$\,AU, $w_2=4$\,AU and $D=20$\,AU. The background pressure gradient has the potential to modify the required radiation pressure (see section \ref{sec:concept}) but we will shortly show that the effect of the thermal pressure gradient is very small, so the exact temperature structure is not too important. This permits us to calculate the radiation pressure force relatively quickly, as we don't have to iteratively run Monte Carlo radiative transfer steps until convergence in the temperature. This is important since we have to trial and modify different dust populations until they yield the observed rotation profile, so many calculations can be required. 
When computing synthetic SEDs from a known dust solution we compute the dust radiative equilibrium temperature using an iterative Monte Carlo radiative transfer scheme.

\section{Results and discussion}

\subsection{Can a thermal pressure gradient explain the rotation curve?}
\label{thermalsec}
We begin by exploring whether a thermal pressure gradient alone can explain the observed azimuthal rotation profile. That is, we consider equation \ref{vphi_wpgrad} with $f_{\textrm{rad}}=0$. Note that we are considering rotation of the mid-plane only at this stage. The density--temperature profile inferred by \cite{2017A&A...601A...5H} (equations \ref{rhoEqun} -- \ref{homanTemp}) gives a mid-plane rotation profile that is always to within 10\,per cent of Keplerian out to 26\,AU and typically closer to within 5\,per cent (the relatively low impact of thermal pressure gradient is also illustrated in Figure \ref{freq}). We hence don't expect the observed sub-Keplerian rotation from the CO fitted {disc} structure alone.

\begin{figure}
	\vspace{-0.33cm}
    \hspace{-20pt}
    \includegraphics[width=9.5cm]{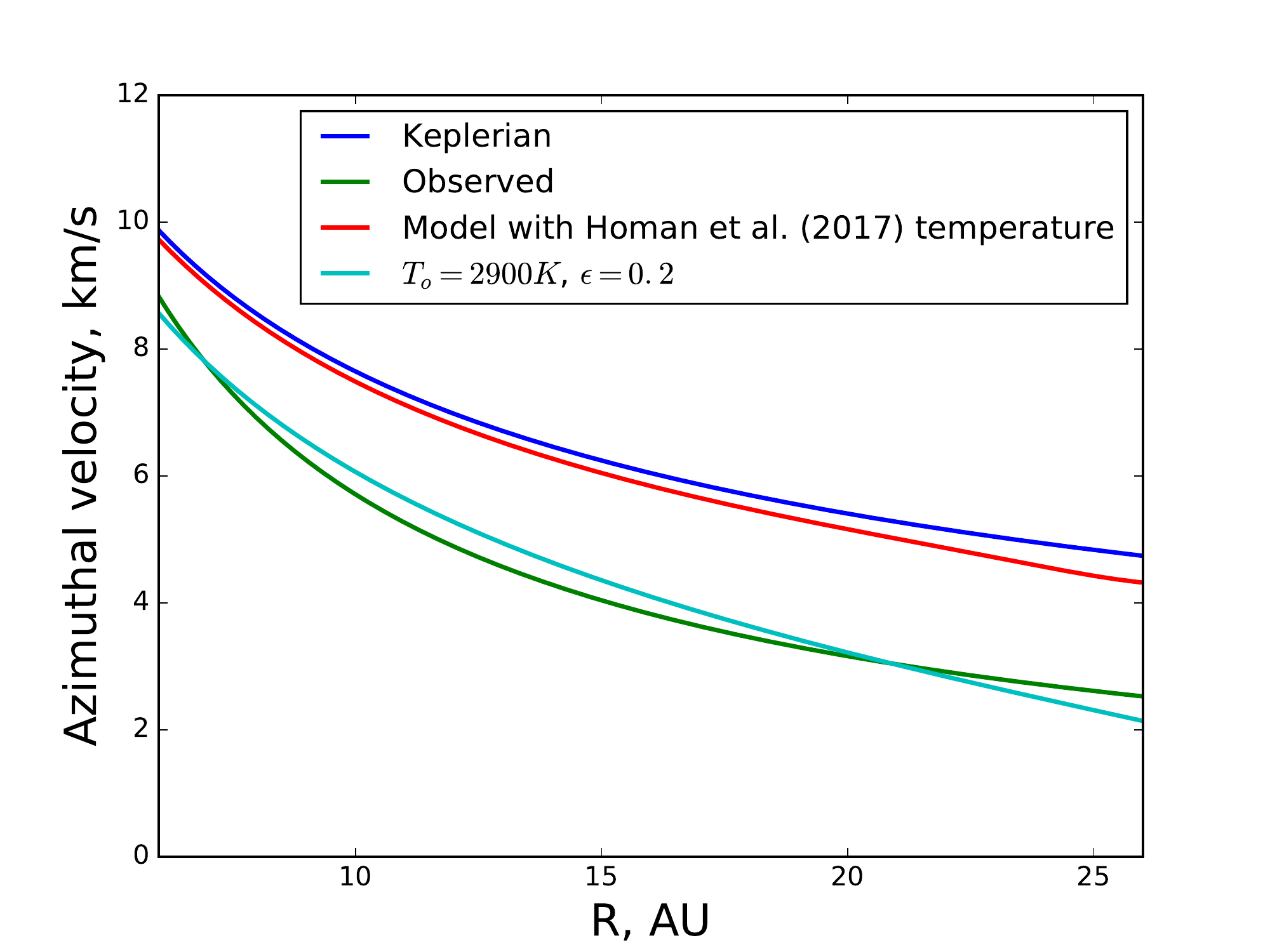}
    \vspace{-0.25cm}
    \caption{The azimuthal velocity as a function of radius without radiation pressure, considering only the impact of a thermal pressure gradient. The {disc} parameters inferred by Homan et al. (2017) do not produce the observed sub-Keplerian rotation. For a power law temperature structure $T=2900(R/AU)^{-0.2}$\,K we get reasonable agreement, but the temperature never drops below 1500\,K in the range considered, which is incompatible with the detection of CO at these radii. }
    \label{vphi_r_thermal}    
\end{figure}

Keeping the density profile the same, we checked for power law temperature profiles ($T=T_o(R/AU)^{-\epsilon}$) that yield azimuthal velocities consistent with the observations, where $T_o$ and $\epsilon$ are free parameters. Solutions are possible, but for temperature profiles that are considerably hotter than observed -- never dropping below $\sim1500$\,K within 26\,AU (e.g. $T_o=2900$\,K, $\epsilon=0.2$ provides a reasonable match -- see Figure \ref{vphi_r_thermal}). Such a thermal structure is incompatible with the observed CO distribution {around} L$_2$ Pup \citep{2017A&A...601A...5H} implying that tweaking of the thermal pressure gradient alone is insufficient to explain the observed azimuthal velocity profile of the gas {around} L$_2$ Pup. 

\subsection{Radiation pressure driven sub-Keplerian rotation}
Using the Monte Carlo radiative transfer scheme discussed in section \ref{nummethod} we searched for dust configurations that resulted in radiation pressure driven sub-Keplerian rotation consistent with that observed, assuming that the background gas pressure profile in the {disc} is that inferred by \cite{2017A&A...601A...5H}. Recall that we consider radially varying, vertically constant, dust populations.

\begin{figure}
	\vspace{-0.33cm}
    \hspace{-20pt}
    \includegraphics[width=9.5cm]{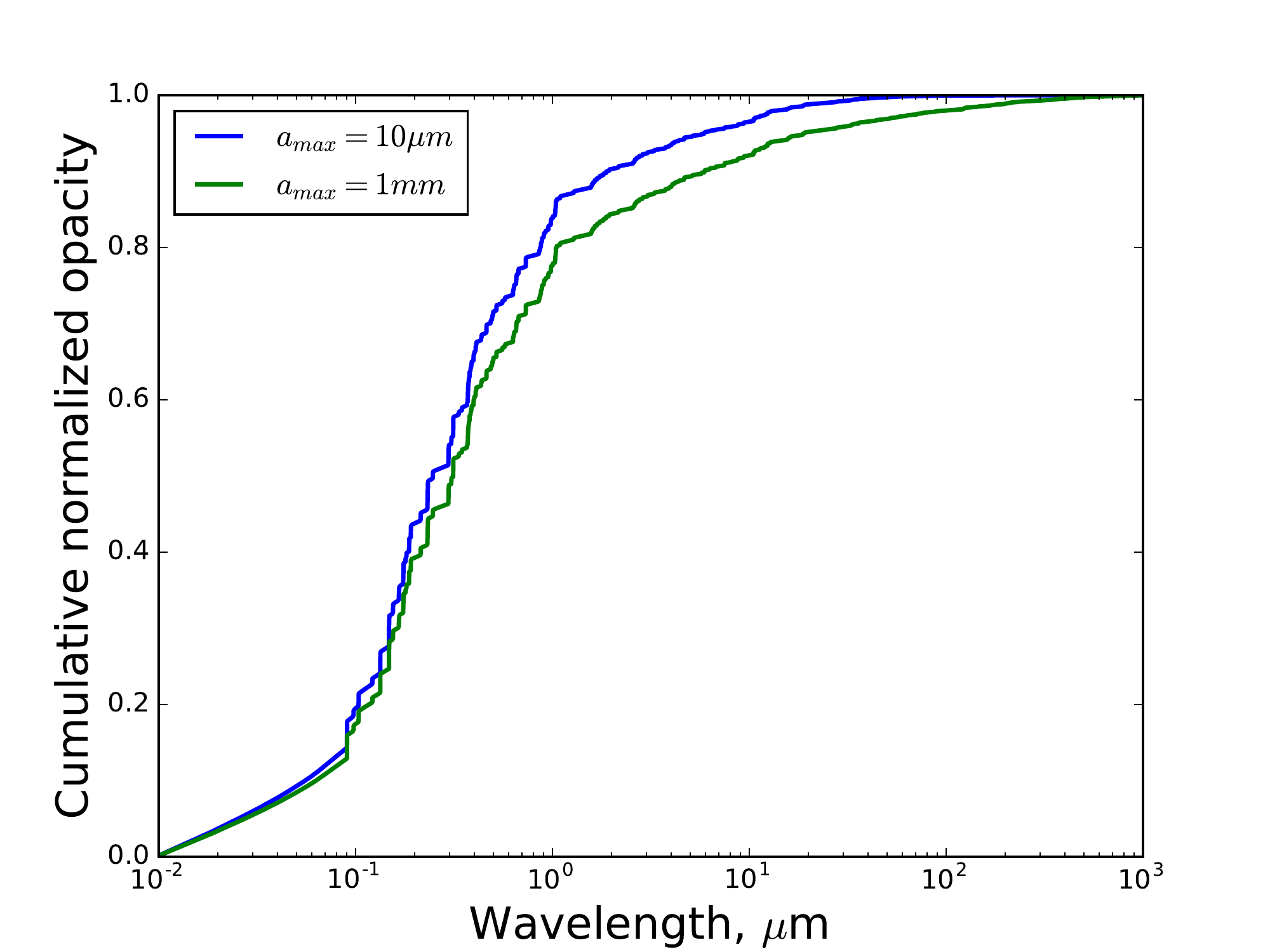}
        \vspace{-0.25cm}
    \caption{The normalized cumulative opacity as a function of wavelength for grain distributions with $q=3.3$, $a_{\textrm{min}}=10^{-3}\,\mu$m and $a_{\textrm{max}}$ of 10$\,\mu$m (red) and 1\,mm (blue)}
    \label{CumulativeOpacity}    
\end{figure}

For a given radiation source, the radiation pressure is sensitive to the dust opacity (see equation \ref{radfequn}) which is influenced by the max/min grain size, dust-to-gas mass ratio, power law distribution and to some extent, the composition. We find that the main parameters are the max grain size and dust-to-gas mass ratio. For simplicity, in the following discussion we generally assume an ISM-like power law of $q=3.3$, but we also explore the effect of $q$ in \ref{sedmodel} (see section \ref{modelparams} for more information of the grain distribution).

The normalized cumulative opacity as a function of wavelength is shown in Figure \ref{CumulativeOpacity} for two grain size distributions that differ only in their maximum grain sizes. The key point here is that the dominant contributor to the opacity is the dust in the size range $0.1-1\,\mu$m. So when varying the dust-to-gas mass ratio or maximum grain size, it is the impact on the population of these grains that affects the azimuthal velocity profile the most. Given this, there is actually a degeneracy between the maximum grain size and dust-to-gas ratio in generating solutions for the rotation profile. Increasing the maximum grain size will deplete the smaller grains somewhat, but increasing the dust-to-gas mass ratio compensates for this. \textit{There is hence a family of possible dust solutions for radiation pressure driven sub-Keplerian rotation {around} L$_2$ Pup.} As we will discuss shortly, these solutions definitely do exist and there are a large number of them, however it is this large number that is unfortunate since it does not permit us to tightly constrain the dust parameters using the rotation curve alone. 

\subsubsection{Compatible dust populations}
\label{sec:compatibledust}
We explore two sets of solutions. In one we choose a fixed dust-to-gas mass ratio and determine the required radial variation of the maximum grain size and in the other we hold the maximum grain size constant and vary the dust-to-gas mass ratio radially. In reality it is likely that both vary radially to some extent, but our approach is more straightforward at this stage. 

\begin{figure}
    \hspace{-10pt}
	\includegraphics[width=9.5cm]{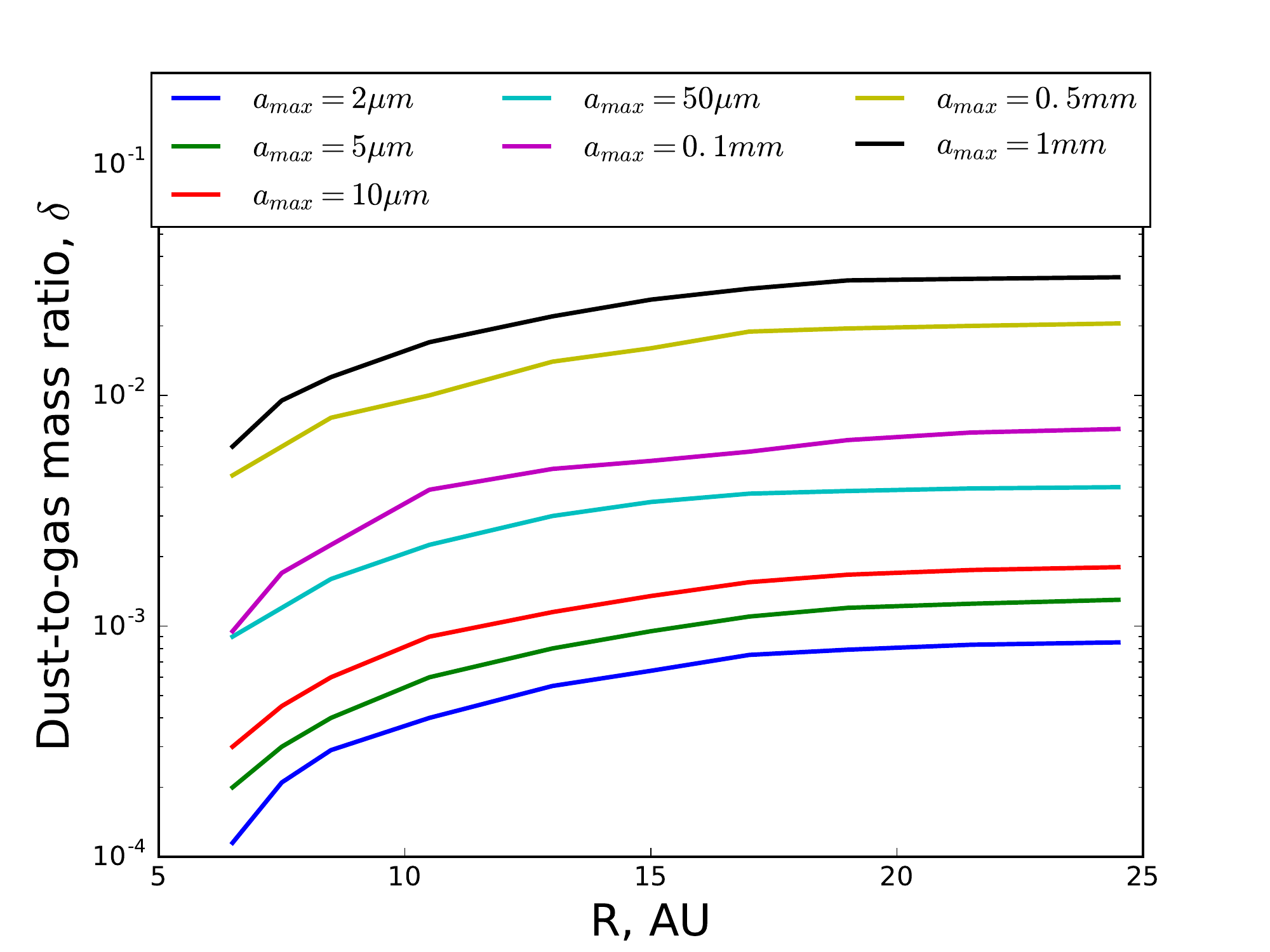}
	\caption{Radial profile of the dust to gas ratio for fixed grain size distributions required to fit the observed azimuthal rotation profile. A larger fixed $a_{\textrm{max}}$ scales up the $\delta$ profile.}
	\label{dgprofile}
\end{figure}

The radial dust-to-gas profiles for different constant $a_{\textrm{max}}$ that satisfy the rotation profile is shown in Figure \ref{dgprofile}. At larger radial distances the opacity has to increase and so the dust-to-gas ratio also has to increase. An ISM-like dust-to-gas ratio of $10^{-2}$, or larger, is achieved for maximum grain sizes $>100\,\mu$m.

\begin{figure}
    \hspace{-10pt}
	\includegraphics[width=9.5cm]{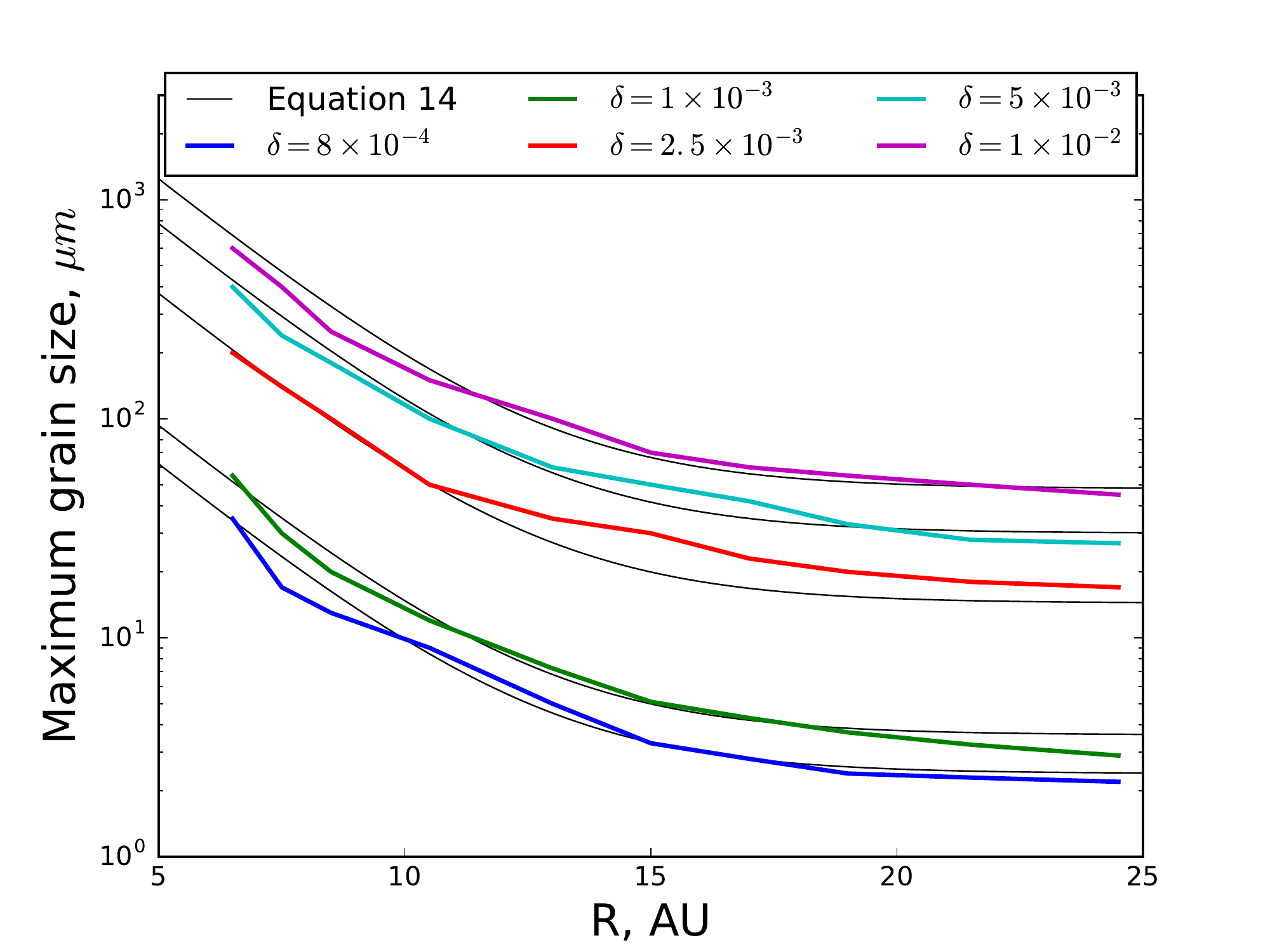}
	\caption{Radial profile of the maximum grain size $a_{\textrm{max}}$ for fixed dust-to-gas mass ratio $\delta$ required to fit the observed azimuthal rotation profile.  A larger fixed $\delta$ scales up the $a_{\textrm{max}}$ profile. Included is an approximate prescription for the radial profile which is given by equation 14 and discussed in section \ref{sedmodel}.}
	\label{amaxprofile}
\end{figure}

Some examples of the radial variation of the maximum grain size for a fixed dust-to-gas mass ratio are shown in Figure \ref{amaxprofile}. As mentioned above the opacity has to increase with radius, which is achieved in the fixed dust-to-gas ratio models by having a decreasing maximum grain size as a function of radius (and hence more grains in the critical 0.1-1$\,\mu$m size range, see Figure \ref{CumulativeOpacity}). We discuss the dust dynamics and grain growth further in sections \ref{dynmodel} and \ref{discussion}, but note here that a radially decreasing maximum grain size could be explained by more rapid grain growth at small orbital distances.

Although there are many possible dust configurations that yield the observed rotation profile, there are some limits. For example we were unable to compute a solution for a dust-to-gas mass ratio of $5\times10^{-4}$ because at large radii in the {disc} we reach a point where the opacity cannot be further increased by reducing the maximum grain size. This is because we enter a regime in which decreasing $a_{\textrm{max}}$ reduces the number of grains in the 0.1-1$\mu$m range and therefore sets an upper limit on the available radiation pressure force. The rotation profile alone can hence offer some further direct constraint on the possible dust properties.

Overall we have found a large range of possible dust parameters that can reproduce the observed rotation curve, owing to the fact that the dust opacity is degenerate. Coupling this family of solutions with other diagnostics will help to further narrow down which subset of the models are valid, which we will do by comparing with the SED in section \ref{sedmodel}. First we check whether our valid dust populations are expected to be dynamically coupled with (and able to exert a back reaction onto) the gas, which would qualify radiation pressure as a viable mechanism for driving sub-Keplerian rotation in L$_2$ Pup.

\subsection{Dust-gas dynamics}
\label{dynmodel}
\begin{figure}
    \hspace{-10pt}
	\includegraphics[width=9.5cm]{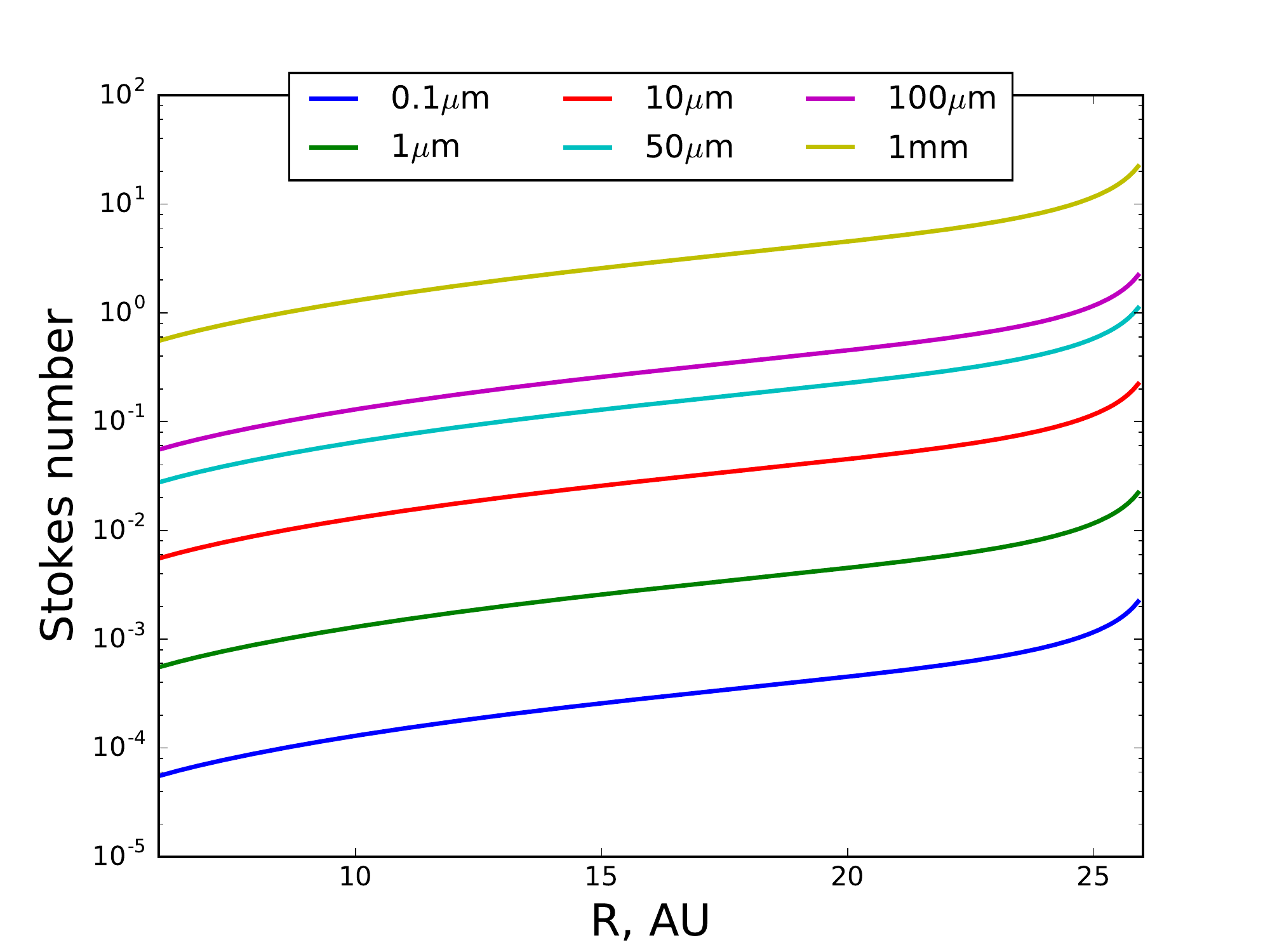}
	\caption{Stokes number of grains as a function of radius {around} L$_2$ Pup. Different lines represent different grain sizes.  }
	\label{stokesfig}
\end{figure}

We estimate the Stokes number of grains in the {disc} in the Epstein regime, that is 
\begin{equation}
    St=\frac{t_sv_{\textrm{kep}}}{R}
\end{equation}
where $t_s$ is the grain stopping time, defined as
\begin{equation}
    t_s=\frac{m_d\rho_g}{K_s\left(\rho_g+\rho_d\right)}
\end{equation}
and the drag coefficient is
\begin{equation}
    K_s=\frac{4}{3}\pi\rho_ga^2v_s\left(1 + \frac{9 \upi \Delta v^2}{128 c_s^2}\right)^{1/2}
\end{equation}
\citep{1975ApJ...198..583K,2006A&A...453.1129P}, where $\rho_g$, $\rho_d$ are the gas and dust volume densities, $m_d$ the grain mass, $a$ the grain size, $c_s$ the sound speed, $\Delta v$ the relative velocity of the dust and the gas and $v_s=\sqrt{8k_bT/(\pi\mu{m_H})}$. The grain mass $m_d$ is computed assuming a density of $3$\,g\,cm$^{-3}$. We plot the Stokes number as a function of radius for various grain sizes in Figure \ref{stokesfig} (setting $\Delta v$ to zero, a larger $\Delta v$ only reduces the Stokes number, increasing the coupling). {The key point from this figure is that the grains in the 0.1-1\,$\mu$m size range (which dominate the opacity for reasonable choices of $q$, Figure \ref{CumulativeOpacity}) have Stokes numbers much less than unity and their azimuthal
motion is therefore expected to be well coupled to that of the gas. Larger grains will be approaching (or exceeding) $St=1$, but since their contribution to the opacity is small the coupling of these grains is less important. However note that in principal large grains could still affect the dynamics if the dust-to-gas ratio was high (approaching unity), due to their inertia. Given the low Stokes number of key grains, radiation pressure on the dust can be responsible for driving the sub-Keplerian rotation observed in the gas. }

\begin{figure}
    \hspace{-10pt}
	\includegraphics[width=9.5cm]{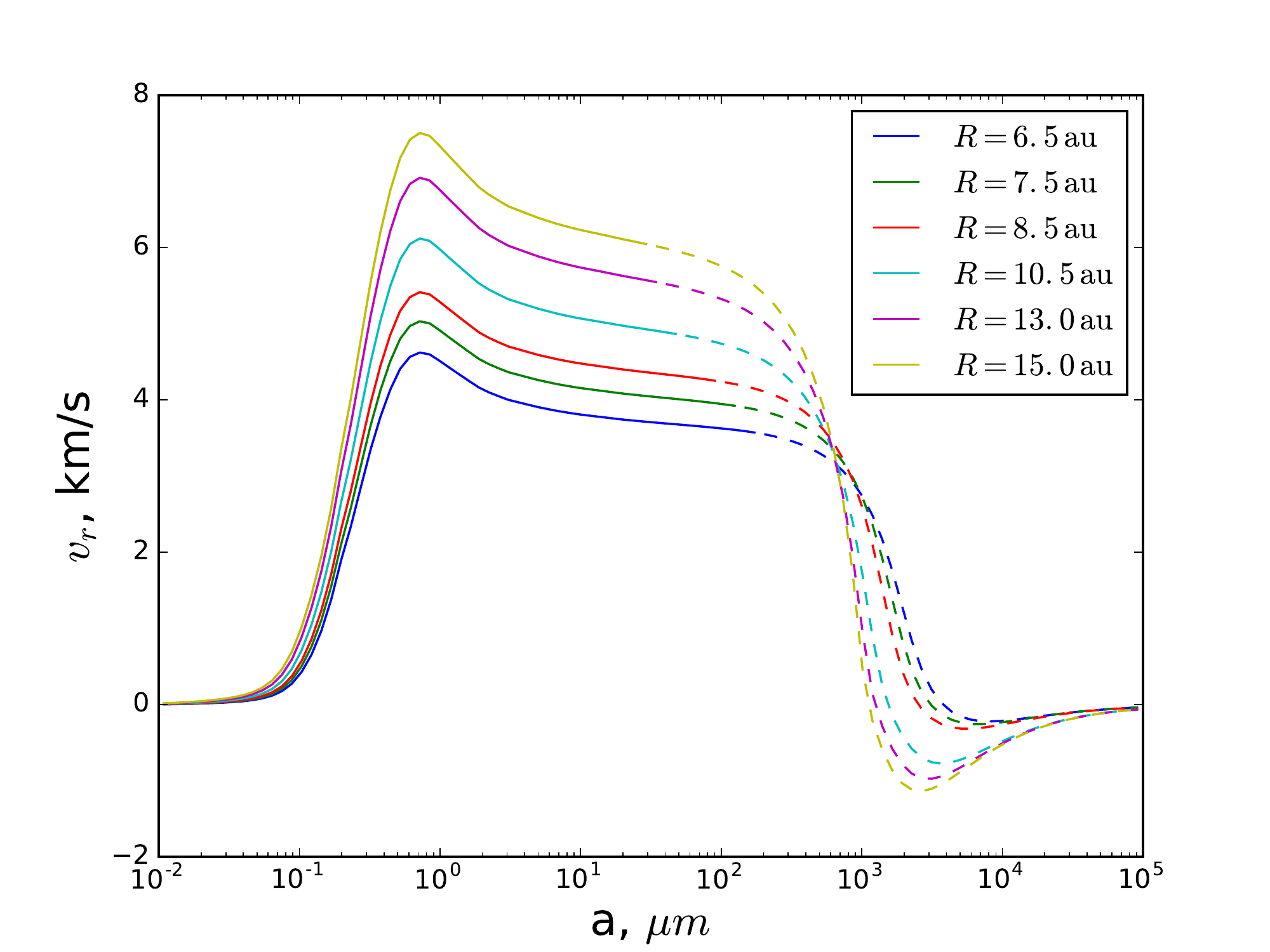}
	\caption{The outward radial velocity of dust grains as a function of grain size at different radii in the disc. This is for a fixed dust-to-gas mass ratio of $2.5\times10^{-3}$. The dashed lines show the radial velocity of grains with sizes above the maximum grain size in the best fit model (assuming negligible contribution to the mass).}
	\label{radial_vel_fig}
\end{figure}

{We have confirmed this by directly solving for the dust and gas velocity at each radius using the coupled equations for dust-gas dynamics  \citep[including the effects of drag, rotation, thermal and radiation pressure][]{1975ApJ...198..583K,1986Icar...67..375N} using a grain-size distribution with a fixed dust-to-gas ratio of $2.5\times10^{-3}$ and radially varying $a_{\textrm{max}}$ that is summarised in Table \ref{fitgrains} (we will show that this distribution fits the SED well in section \ref{sedmodel}).} We find that the differences between the gas and dust azithumal velocity is typically less than one per cent, confirming the tight coupling. Interestingly, although the gas radial velocity remains small ($\ll 1\,{\rm km\,s}^{-1}$, with the radial drag force balanced by the coriolis force), the large radiation pressure drives the dust to large outward radial velocities at all sizes $\gtrsim 0.1\,\umu$m (Figure \ref{radial_vel_fig}). Only for  sizes above $0.1$--$1\,{\rm cm}$ does gravity overcome radiation pressure, allowing the grains to remain in the {disc} and be re-accreted by the star. For the smallest grains ($<0.1\,\mu$m) the coupling with the gas is so tight that the outward velocity is also low. 

Note that the situation is analogous to the familiar case of radial drift of dust in protoplanetary discs except that in this case
the relative motion of the dust and gas is driven by radiation
pressure on the dust rather than the effect of radial pressure
gradients on the gas. In both cases it can readily be shown from
considering the balance of drag and Coriolis force in the azimuthal
direction that the dust-gas relative velocity in the azimuthal
direction is a factor $St$ times the relative velocity in the radial direction. This is
in line with our finding here that the dust-gas relative velocity is significant in the radial direction while the two fluids are tightly coupled in the azimuthal direction.

The large radial velocity in dust means that the {disc} will be depleted of dust within $\sim 20\,{\rm yr}$, which suggests that the {disc} must be replenished on short time-scales, we will discuss this replenishment further in section \ref{discussion}. This velocity of the grains also implies that we should ideally re-compute the grain surface density profile to be consistent with the velocity structure, which would require  multiple iterations and both a variable dust-to-gas ratio and graiz size mixture at different radii. Given that there are other approximations (e.g. vertically constant grain properties) for simplicity we leave such considerations for future work.

In this and the previous section we have demonstrated that there are dust configurations that result in opacities sufficient to drive sub-Keplerian rotation consistent with that observed, that are dynamically coupled to the gas. We also showed in section \ref{thermalsec} that a thermal pressure gradient alone cannot be responsible. We hence conclude that radiation pressure is a capable and necessary mechanism to drive the observed rotation profile of matter in the {disc} {around} L$_2$ Pup.

\subsection{SED modelling}
\label{sedmodel}
We have now shown that radiation pressure can theoretically explain the sub-Keplerian rotation {around} L$_2$ Pup. We have already discussed that the rotation profile offers only a weak constraint on the dust given the degeneracy between the maximum grain size and dust-to-gas mass ratio. However even a weak additional constraint might still help to yield an improved insight into the dust {around} an AGB star. We recomputed the dust radiative equilibrium temperature for our dust solutions and generated synthetic SEDs, which we compare with the observed NACO/VLT, VLTI/MIDI, and other data, summarised in \cite{2014A&A...564A..88K}.

\begin{figure}
    \hspace{-10pt}
	\includegraphics[width=9.5cm]{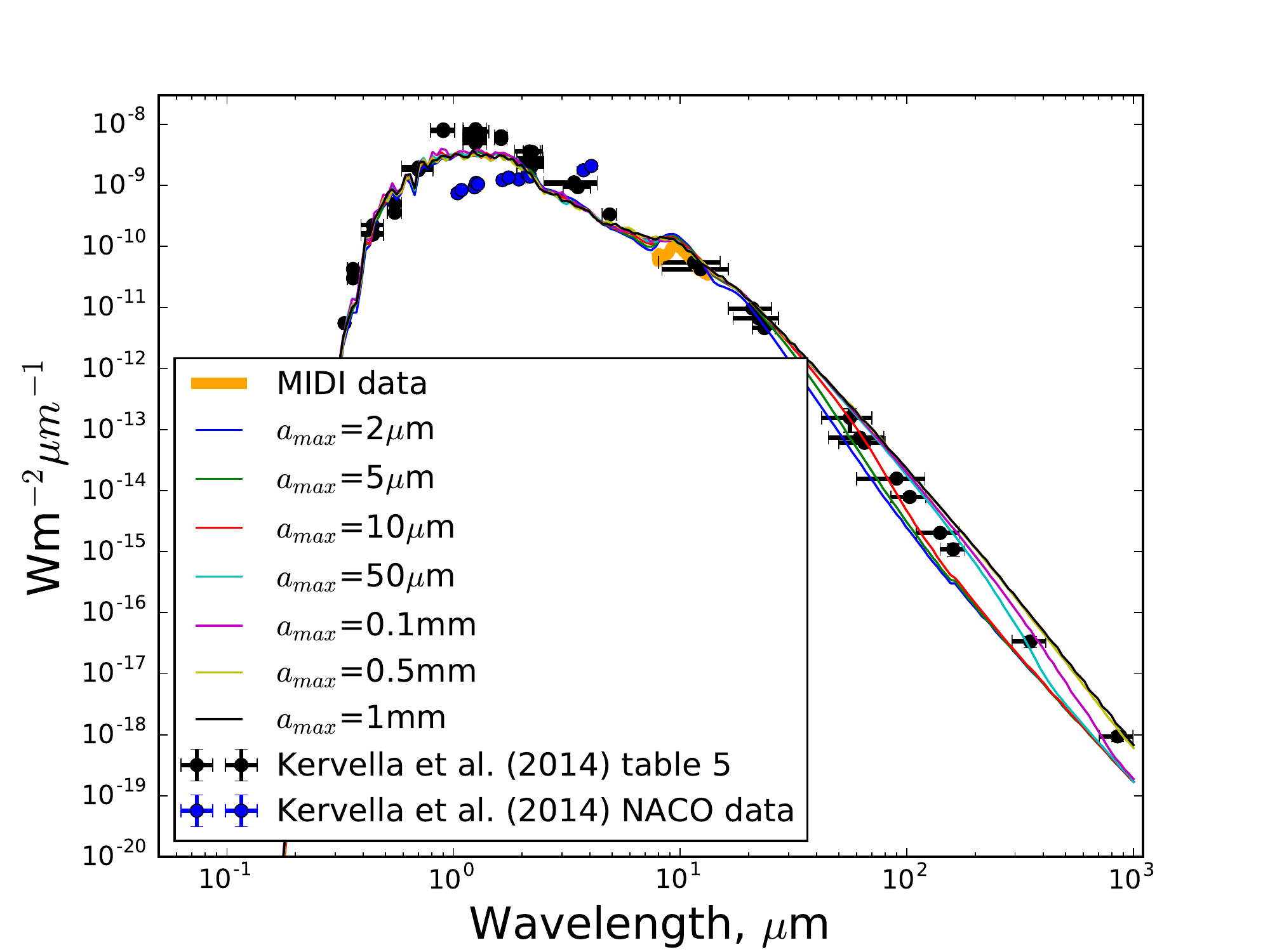}
	
    \hspace{-10pt}	
	\includegraphics[width=9.5cm]{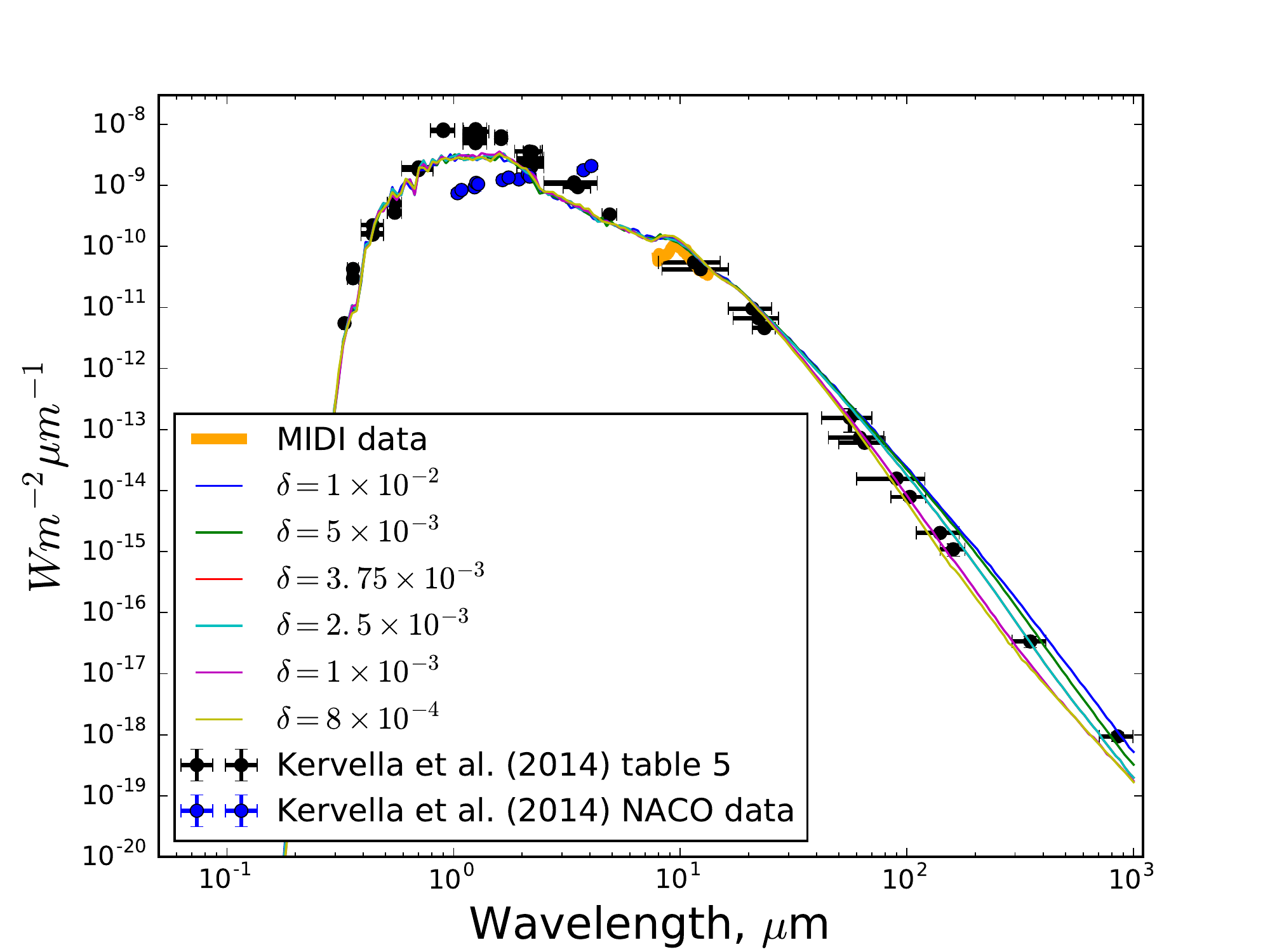}
	\caption{A summary of our model SEDs. The upper panel is for models with fixed maximum grain size $a_{\textrm{max}}$ and radially varying dust-to-gas ratio $\delta$. The lower panel is for fixed dust-to-gas ratio and radially varying maximum grain size. }
	\label{sed_allmodels}
\end{figure}

Figure \ref{sed_allmodels} shows the total SED for our models, with the upper panel showing the results for models with fixed maximum grain size and the lower panel showing the results for models with fixed dust-to-gas mass ratio. 

\begin{figure}
    \hspace{-10pt}
	\includegraphics[width=9.5cm]{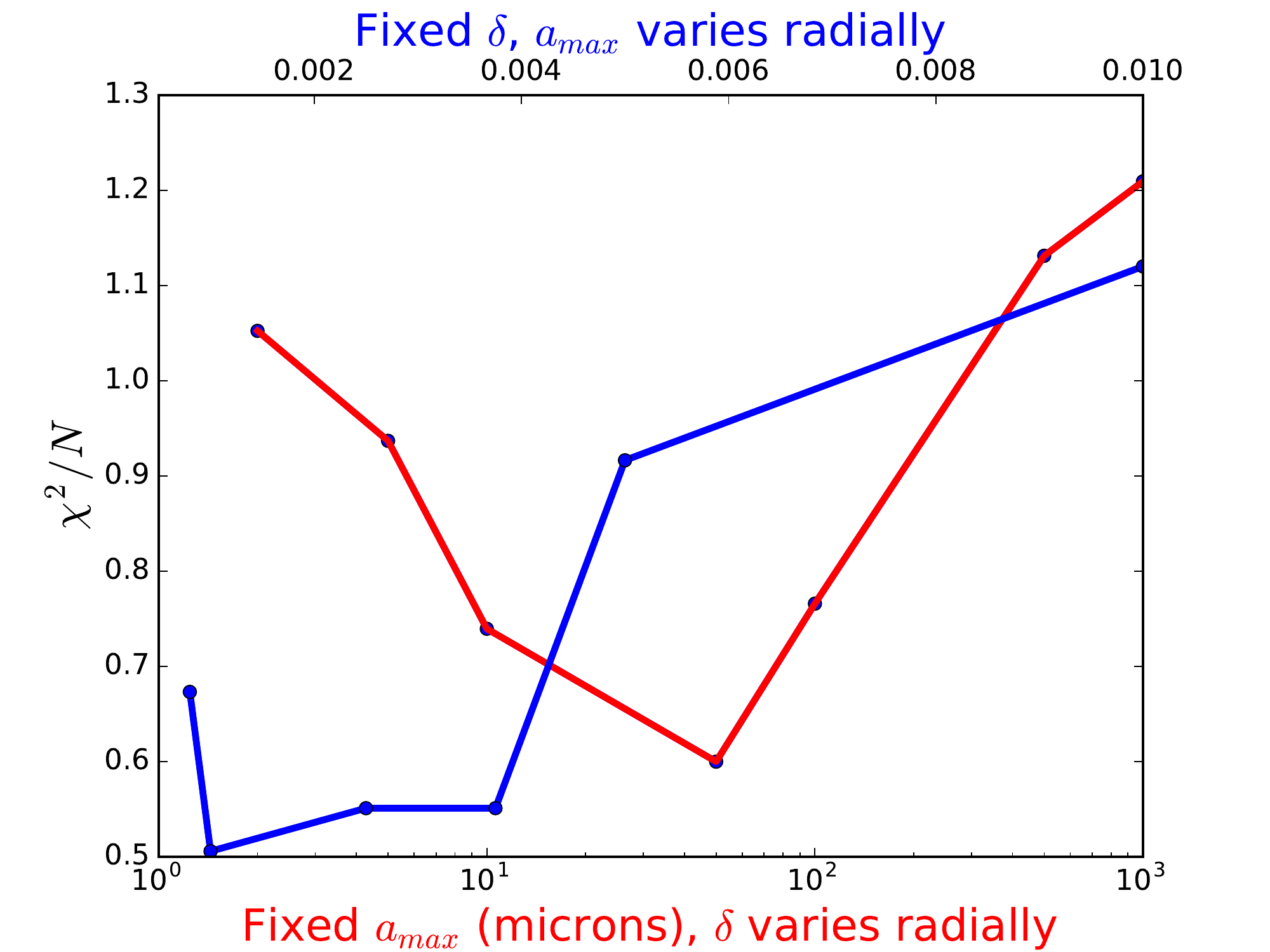}
	\caption{A measure of the chi-square goodness of fit of our model SEDs compared to the observed data. }
	\label{chisquare}
\end{figure}

We evaluate {the goodness of fit of} each model using a chi-square measure for the $N$ points of the observed SED longward of $10\,\mu$m, where dust dominates the emission (shortward of $10\,\mu$m the SED is dominated by the stellar contribution, as we will discuss below). For the observed data the filter width dominates over the flux uncertainty. This, coupled with the fact that the SED is a single valued function beyond $10\,\mu$m, permits us to compare the observed and synthetic \textit{wavelengths} at a given flux in our $\chi^2$ measure
\begin{equation}
    \chi^2=\frac{1}{N}\sum\frac{\left(\lambda_{\textrm{{obs}}}-\lambda_{\textrm{sim}}\right)^2}{\Delta\lambda_{obs}^2}.
\end{equation}
{The result of this comparison is shown in Figure \ref{chisquare}.  Models with fixed dust-to-gas ratio that vary the maximum grain size radially do the best overall job for dust-to-gas mass ratios in the range $1-4\times10^{-3}$. The $\delta=10^{-3}$ model has the best goodness of fit measure, but is not consistent with the $\sim$mm observations within uncertainties. Conversely the model with $\delta=2.5\times10^{-3}$ is the only one that is consistent with all observed points longward of $10\,\mu$m within uncertainties -- we hence refer to this model as our best match.}

{For models with fixed maximum grain size and and radial variation of dust-to-gas ratio the best solutions are those with  $a_{\textrm{max}}\sim50\,\mu$m. However, varying the maximum grain size does not give as good a match as models that vary the dust-to-gas mass ratio.}

\begin{figure*}
    \hspace{-10pt}
	\includegraphics[width=13cm]{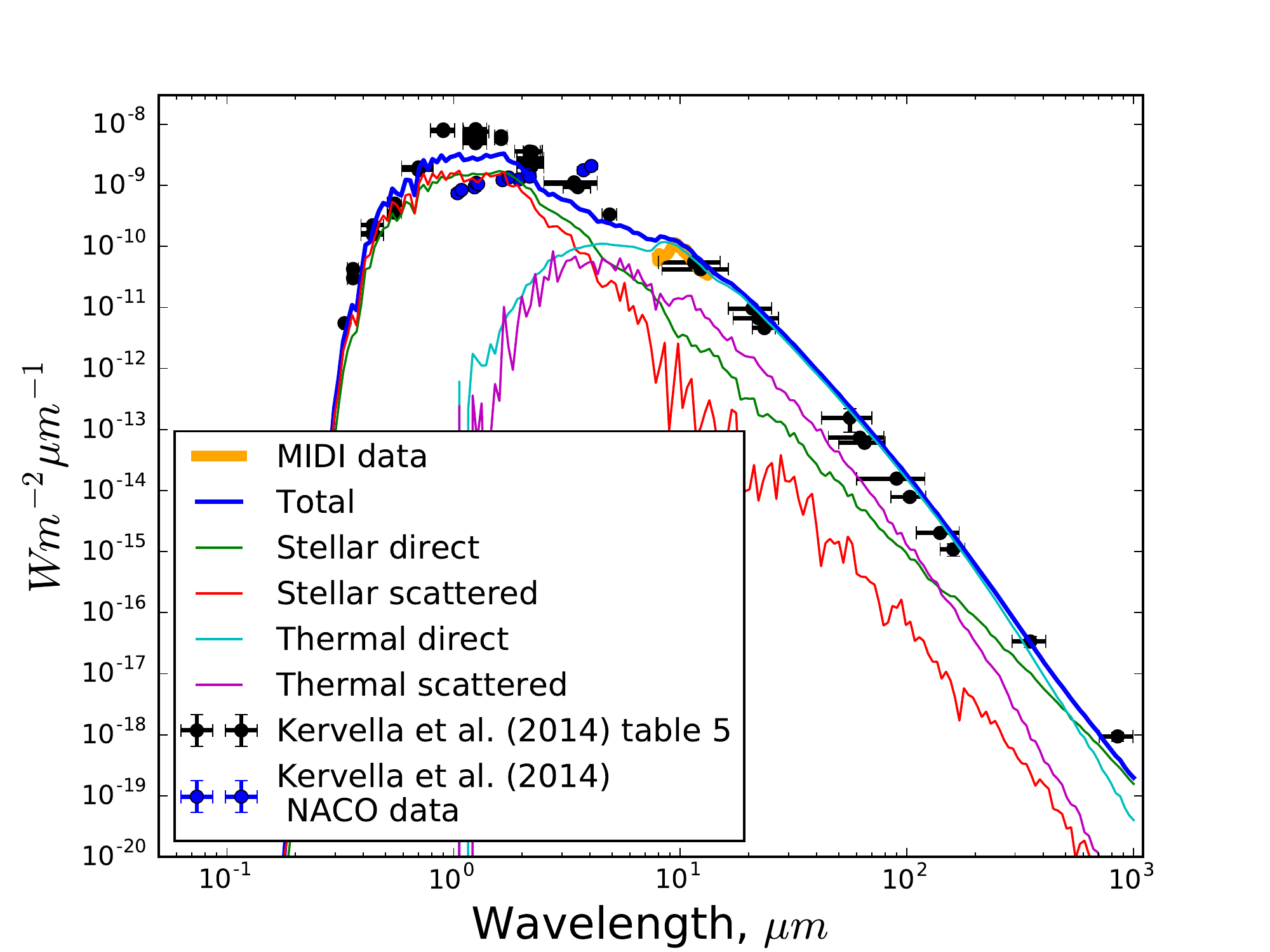}
	\caption{The SED of the model with fixed $\delta=2.5\times10^{-3}$, our model that best matches the data, including the separate photon source contributions. The points from $10\,\mu$m longwards are those affected by the dust distribution. This best fit model lies within the uncertainties of all observed data points longwards of $10\,\mu$m.  }
	\label{bestfitsed}
\end{figure*}

Figure \ref{bestfitsed} shows the SED for the $\delta=2.5\times10^{-3}$ model, decomposed into its component parts: direct and scattered stellar photons, and direct and scattered thermal (dust continuum) photons. 
As mentioned above Figure \ref{bestfitsed} confirms that shortward of $10\,\mu$m the SED is dominated by the stellar contribution and we hence see only negligible differences with different dust models. Note that the point at 1\,mm is actually primarily set by the stellar spectrum, which is in the blackboday regime by this wavelength (see section \ref{sec:source}), but does require a small amount of large grains to boost the flux to the observed value. \\

 Recall that Figure \ref{amaxprofile} shows the radial variation of grain sizes for fixed dust-to-gas ratio. As a convenience we find that this can be approximately described by
 \begin{equation}
     a_{\textrm{max}}(R)\approx a_{\textrm{max}}(6\,\textrm{AU}) \\
      \left\{6\times10^{-2}+12\exp\left[-2.5\left(\frac{R}{6\,\textrm{AU}}\right)\right]\right\}\,\mu{m}.
      \label{fitequn}
 \end{equation}
 This approximation is compared against our results in Figure \ref{amaxprofile}.  For future reference, the dust properties in all bins of the best two models ($\delta=10^{-3}, 2.5\times10^{-3}$) are summarised in Table \ref{fitgrains}. We re-emphasize that there will be other solutions too, and that in reality both the dust-to-gas ratio and $a_{\textrm{max}}$ will vary radially.

\begin{table}
    \caption{A summary of the grain parameters in our models that give the best matches to both the rotation profile and SED of L$_2$ Pup. These models have a power law distribution $q=3.3$, a minimum grain size of 1\,nm and a dust-to-gas mass ratio of $\delta=1\times10^{-3}$ and $\delta=2.5\times10^{-3}$. Note that there will be other possible solutions for different $q$, as well as in scenarios where both $\delta$ and $a_{\textrm{max}}$ can vary radially. }
    \label{fitgrains}
    \begin{tabular}{c|c|c|c}
        \hline
        $R_{\textrm{min}}$ & $R_{\textrm{max}}$ & $a_{\textrm{max}} (\delta=2.5\times10^{-3})$ & $a_{\textrm{max}} (\delta=1\times10^{-3})$   \\
        (AU) & (AU) &  ($\mu$m) & ($\mu$m) \\
        \hline
        6 & 7 &  200 & 55\\
        7 & 8 &  140  & 30 \\    
        8 & 9 &  100  & 20\\  
        9 & 12 & 50 & 12\\  
        12 & 14 &  35 & 7.25\\          
        14 & 16 &  30  & 5.1\\
        16 & 18 &  23 & 4.3\\    
        18 & 20 &  20 & 3.7\\          
        20 & 23 &  18 & 3.25\\        
        23 & 26 &  17 & 2.9\\         
        \hline
    \end{tabular}
\end{table}

There have been other studies that have modelled the SED of L$_2$ Pup \citep[e.g.][]{2014A&A...564A..88K, 2016MNRAS.460.4182C}, but no other model in the literature provides such a good simultaneous fit to so much of the SED. In particular, fitting beyond 10\,$\mu$m has not been so successful in the past. There are hence members of the family of dust solutions that satisfy the observed rotation profile that simultaneously reproduce the SED.

We also modified one of our best models, with fixed $\delta=10^{-3}$, to probe the impact of the grain power law $q$. Our models throughout have assumed $q=3.3$, but we ran three additional calculations with $q=3$, $q=3.5$ and $q=4.2$ \citep[the latter derived by][for carbon rich winds]{2013pccd.book.....G} for fixed $\delta=10^{-3}$. Changing $q$ affects the slope of the jump in cumulative opacity seen in Figure \ref{CumulativeOpacity}. We found that we were able to locate solutions to the rotation profile for $q=3, 3.5$ by simply changing the maximum grain size. In the case of $q=4.2$ we were only able to obtain solutions by also modifying the minimum grain size/dust-to-gas ratio, since the opacity was always too high with our fiducial parameters. For the lower $q$ models that still give solutions, the SED $\chi^2$ value doesn't change by more that 1.5\,per cent over $q=3$ to $q=3.5$ for fixed $\delta=10^{-3}$. The value of $q$ is therefore of secondary importance, at least over the range considered here. The $q=4.2$ SED does still give  good agreement beyond 10\,$\mu$m, but interestingly causes the stellar scattered light flux shortward of 1\,$\mu$m to deviate below the observed values.

\subsection{Further discussion}
\label{discussion}

\subsubsection{On dust replenishment}
In section \ref{dynmodel} we found that grains in the size range $\sim0.1\,\mu\textrm{m}-0.25\,\textrm{cm}$ are rapidly blown out of the {disc} by the intense radiation pressure, giving a depletion timescale of only tens of years. Given that the probability that the observed {disc} {around} L$_2$ Pup is only of such an age is incredibly low, there must either be proportionally strong dust formation and growth to replenish the population, or some additional mechanism hindering the outward migration. We reiterate that although there is rapid outward radial motion of the dust, the azimuthal dust-gas coupling is actually very tight and the gas does not move radially with any significant velocity. 

Our models in this paper are dynamically and geometrically quite simple, making a robust estimate of the mass loss rate in dust difficult. As a zeroth order estimate the dust mass loss rate is simply the dust mass times the $\sim20$\,AU of the dusty {disc} divided by the clearing velocity. However this neglects the fact that the {disc} is highly turbulent ($\sim1$\,km\,s$^{-1}$), which will hinder the outward radial propagation of the dust. Processes such as dredging and shearing instabilities may also hinder the radial dust propagation. Nevertheless, we make the above estimate for our models as follows. 

The total {disc} mass is $2.2\times10^{-4}$\,M$_{\odot}$ \citep{2017A&A...601A...5H}, however only 30\,per cent of this is in the range 6--26\,AU, with the majority from 2--6 AU. We hence consider a {disc} mass of $6.6\times10^{-5}$\,M$_\odot$, dust-to-gas ratio of $10^{-3}$ and a clearing time of 20 years, which yields a required replenishment rate of $3.3\times10^{-9}$\,M$_\odot$\,yr$^{-1}$. Such a rate is completely feasible for AGB stars in the solar neighbourhood, where dust mass loss rates are typically in the range $10^{-9}-10^{-7}$\,M$_\odot$\,yr$^{-1}$ \citep[e.g.][Trejo et al. in prep.]{1989ApJ...341..359J,2015ASPC..497..405T}. 

There are existing, somewhat uncertain, estimates of the dust mass loss rate from L$_2$ pup. \cite{2002MNRAS.337...79B} estimated a dust mass loss rate of L$_2$ Pup of $5\times10^{-10}$\,M$_\odot$\,yr$^{-1}$, assuming a velocity of 2.5\,km\,s$^{-1}$, dust-to-gas ratio of $\sim10^{-3}$. The difference in their assumed velocity with our typical $5$\,km\,s$^{-1}$ gives a factor 2 increase in the dust mass loss rate and hence a deficiency of factor 3.3 between their mass loss rate and our required rate. Furthermore, \cite{2002ApJ...569..964J} estimated a dust mass loss rate of $1.9\times10^{-9}$\,M$_\odot$\,yr$^{-1}$ for a 3.5\,km\,s$^{-1}$ wind, which translating to a 5\,km\,s$^{-1}$ wind gives a replenishment rate of $2.7\times10^{-9}$\,M$_\odot$\,yr$^{-1}$, which is very close to our required value. Overall then, observationally inferred mass loss rates are somewhat consistent with (albeit a bit lower than) the value required from our models. 

The slightly higher replenishment rate in the models is easily accounted for by uncertainties in the model and observations. Uncertainties in the {disc} mass alone \citep[for which][quote the lower limit as a factor 3.4 smaller than the value we consider for our calculation]{2017A&A...601A...5H}  can account for the discrepancy. This is also without any inclusion of uncertainty in the CO/H ratio when calculating the {disc} mass, which they assumed to be $10^{-4}$ \citep[][find a higher CO/H ratio at small radii, which would decrease the {disc} mass, though the value of this ratio is highly uncertain]{1988ApJ...328..797M}. In addition to this there are processes such as turbulence that will slow the outward propagation of grains. Furthermore \cite{2008A&A...487..645R} placed a lower limit in uncertainty of $\sim 3$ on observationally inferred mass loss rates at the time of their work.  

Overall then, the high velocities in grains predicted by our models are not incompatible with either the kinematic observations of gas (since the dust and gas are only azimuthally coupled) nor the required replenishment rate (within uncertainties). Reducing uncertainties with future observations will help to constrain our models further and confirm whether the dust population can indeed be sustained. 

As a final comment, one might speculate that variability on the timescale of the dust production rate may result in a corresponding variability of the {disc} structure as dust depletes, the opacity drops and the azimuthal velocity becomes more Keplerian (or vice versa). This could be surveyed observationally.

\subsubsection{On the possibility that the circumstellar matter is more wind-like than disc-like}
The analysis of \cite{2016A&A...596A..92K} found a  $r^{-0.853\pm0.059}$ azimuthal velocity scaling, which is close to the $r^{-1}$ scaling expected for a slow wind that conserves angular momentum. This coupled with our high outward radial velocities of large grains raises the possibility that the circumstellar outflow might be more like a wind than a disc. However, comparing the thermal pressure gradient, centrifugal and gravitational forces (e.g. equation \ref{vphi_wpgrad} without radiation pressure) we find that the gas acceleration is actually slightly inwards radially, inconsistent with a strong wind radially outwards. Furthermore, no evidence for a fast wind was found in the observational kinematic study of \cite{2016A&A...596A..92K}. We therefore conclude that the {disc} interpretation is the more applicable and that there is tight azimuthal coupling between dust and gas, but only weak radial coupling.

\subsubsection{Sensitivity to grain composition}
\label{grainComposition}
In addition to pure silicates, we also computed additional models using iron rich and iron deficient magnesium-iron silicates (see section \ref{modelparams}). Figure 1 of \cite{2006A&A...460L...9W} shows that we expect higher near-infrared absorption efficiency for small ($<<1\,\mu$m) iron rich grains than iron poor, which makes such grains more effective at driving a radiation pressure induced wind. However \cite{2008A&A...491L...1H} showed that the larger scattering efficiency of micron sized  grains is still high enough to permit iron-free grains to drive a wind. 

\begin{figure}
	\includegraphics[width=9cm]{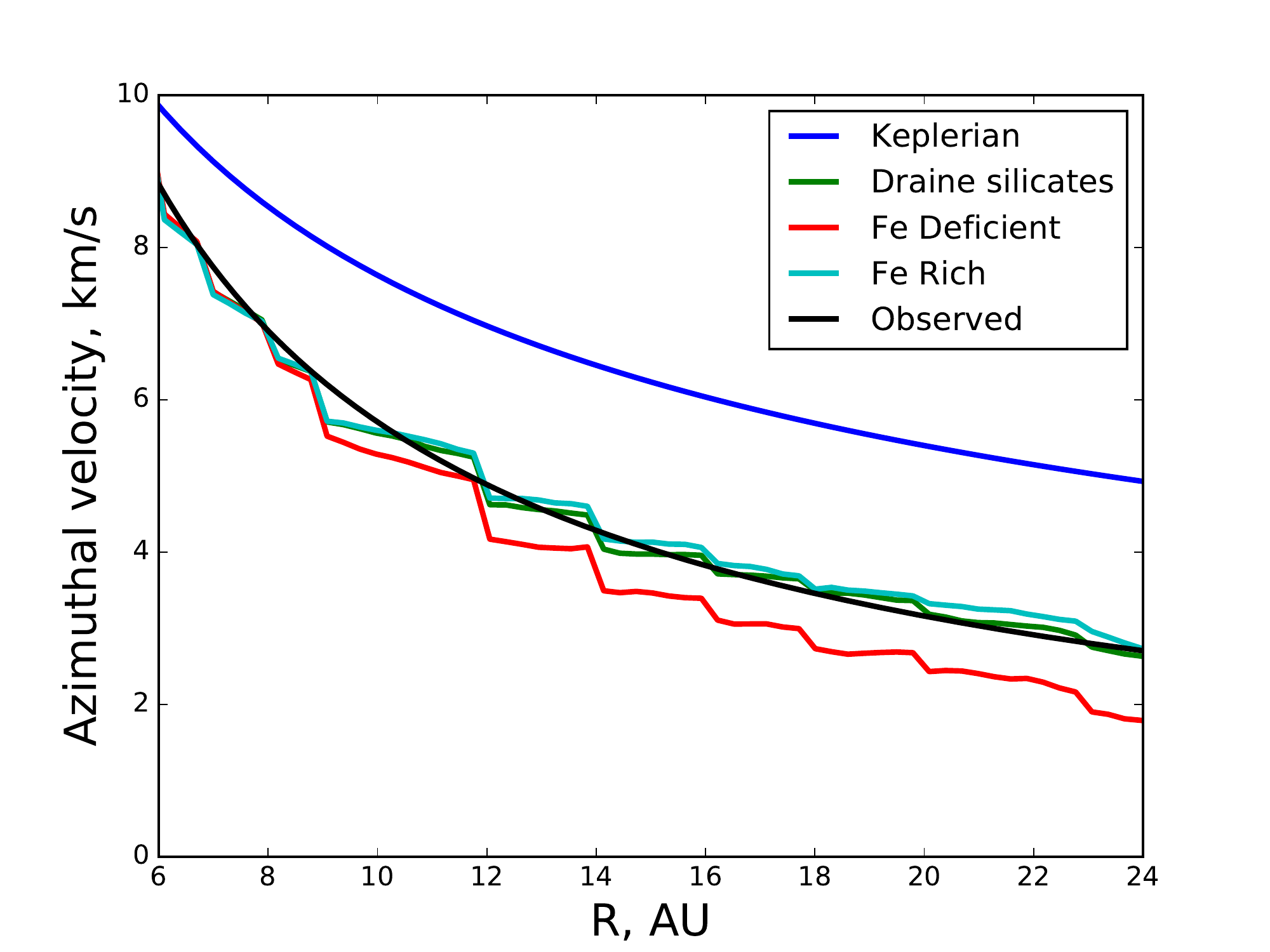}
	\caption{A comparison of the rotation profile for grain populations that differ only in their optical constants (the grain density is forced to be the same in each case). The black line is the observed rotation profile.  }
	\label{composition_rotvel}
\end{figure}

\begin{figure}
	\includegraphics[width=9cm]{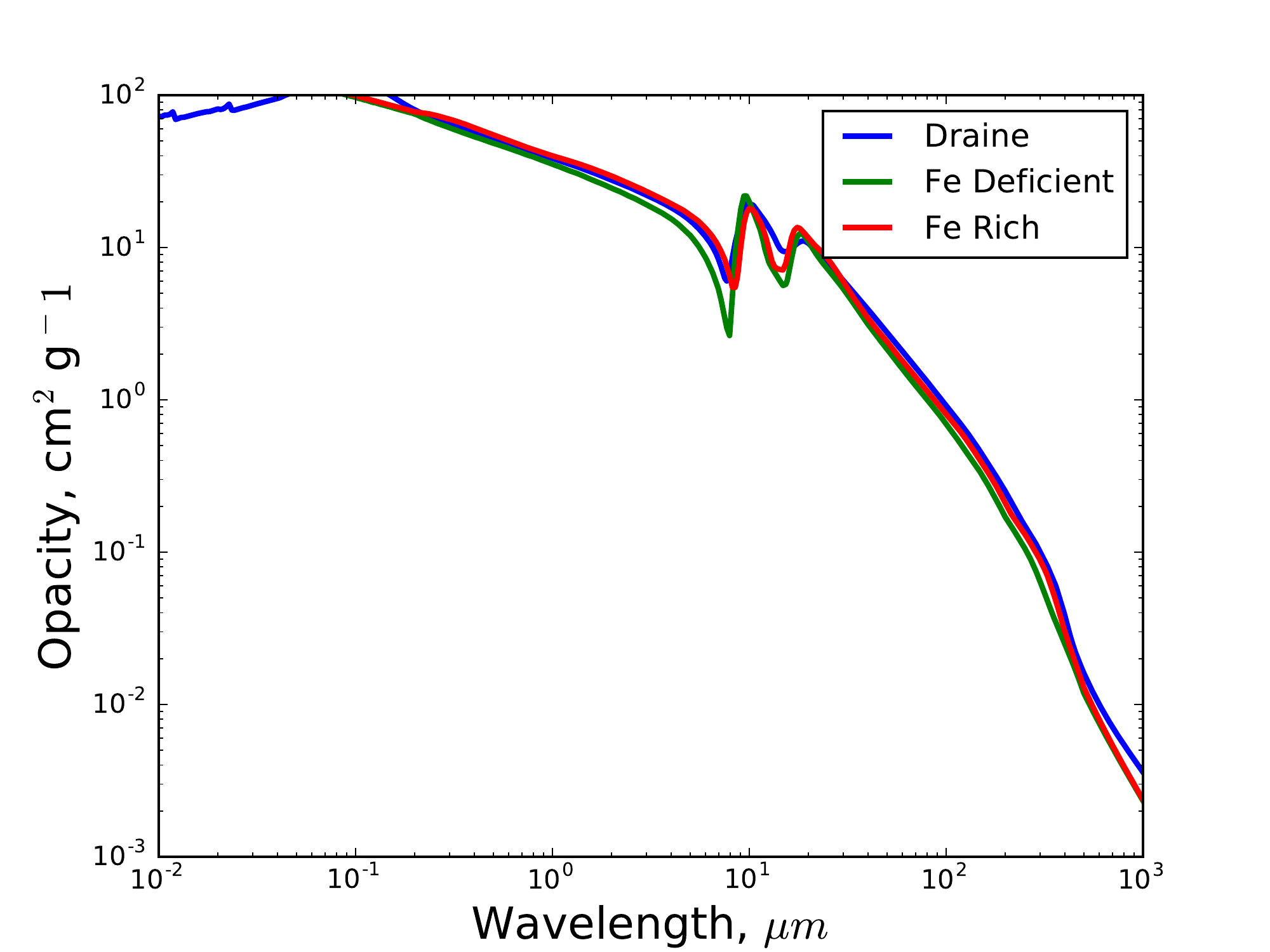}
	\includegraphics[width=9cm]{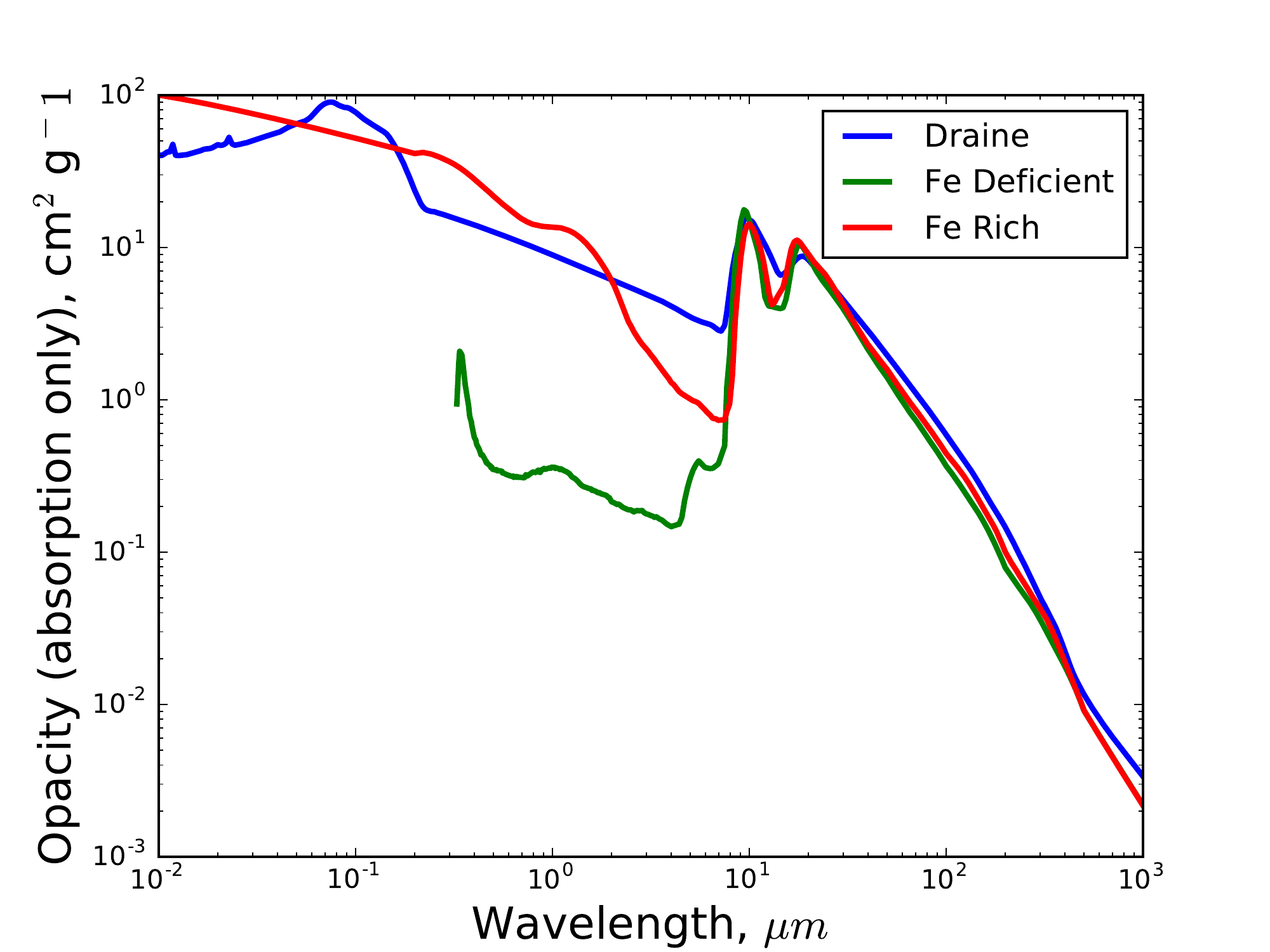}
	\includegraphics[width=9cm]{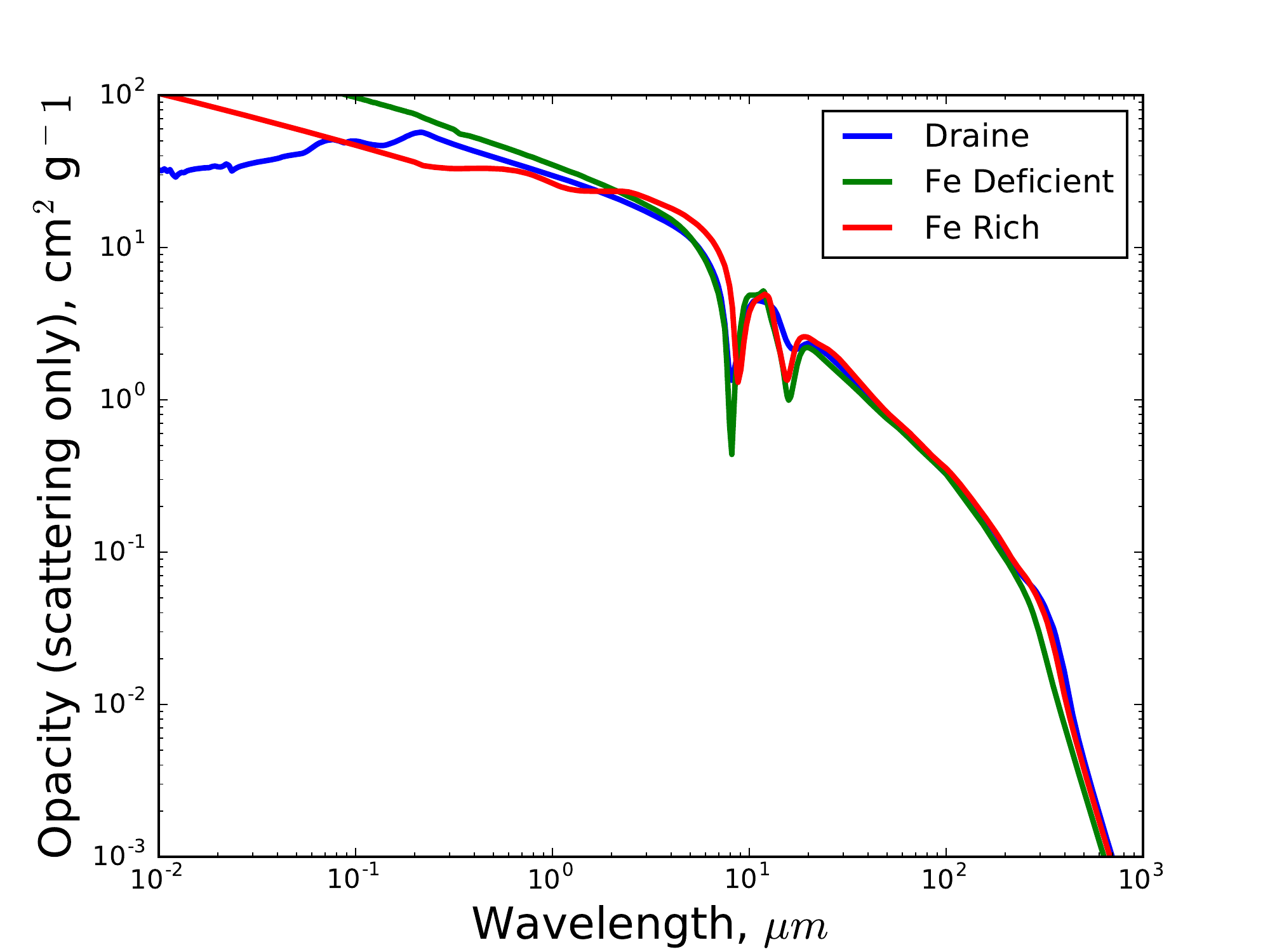}	
	\caption{The opacity breakdown of different grain types. The top panel is the total opacity, the middle the absorption only and the bottom is scattering opacity only. Note that the iron deficient absorption opacity drops to $\sim10^{-4}$\,cm$^{2}$\,g$^{-1}$ at less than 0.3\,$\mu$m}
	\label{composition_opacities}
\end{figure}

To simplify our initial comparison we force each grain type to have the same density (3.5\,g\,cm$^{-3}$), so they only differ in their optical constants. We assume the same radial profile of grains ($a_{\textrm{min}}$, $a_{\textrm{max}}$, $q$), which is based on the solution for \cite{2003ApJ...598.1017D} silicates and $\delta=10^{-3}$ (see Table \ref{fitgrains} for the grain sizes). The rotation curves for the different compositions are shown in Figure \ref{composition_rotvel}. Iron rich and \cite{2003ApJ...598.1017D} silicates show similar profiles, but the iron deficient grains show a slower rotation curve, implying that the product of the flux and the opacity is higher (equation \ref{radfequn}).  To understand this, Figure \ref{composition_opacities} compares the absorption, scattering and total (absorption plus scattering) opacities of each grain type. Although the absorption opacity of iron rich grains is indeed higher, this is compensated for by the micron sized grains being able to efficiently scatter photons \citep{2008A&A...491L...1H}. The total opacity is therefore similar in each case. The implication of this is that in the iron deficient case there is higher flux at large radii (since it has been scattered rather than absorbed). We verified this by checking the radial mid-plane flux profile and it is indeed higher in the iron deficient case. For grain distributions that differ only in their optical constants, iron deficient grains are hence more capable of driving sub-Keplerian rotation, particularly at larger distances.

In addition to our checks on grains that only differ in their optical constants we also computed models with the appropriate grain densities (2.74, 3.71 and 3.5\,g\,cm$^{-3}$ for iron deficient, iron rich and \cite{2003ApJ...598.1017D} silicates respectively) and for which the dust-to-gas ratio is scaled to give the observed rotation profile. We find that iron rich grains and \cite{2003ApJ...598.1017D} silicates both give similar results for the same dust-to-gas mass ratio, including for the SED and outward radial acceleration of grains. However, iron deficient grains require a $\sim$40\,per cent lower dust mass in order to reproduce the rotation profile and the outward radial grain velocities are $\sim1\,$km/s (20\,per cent) faster. This combination means that iron deficient grains require a slightly lower dust production rate sustain their population. However, the SED of the iron deficient grains in this case is not good, significantly underestimating the flux of the points from $3-30\,\mu$m.

It is important to note that for different grain size distributions (larger $a_{\textrm{max}}$) iron deficient grains can still yield the rotation profile using the same dust mass as iron rich grains and, in such a case, the SED can also provide a reasonable match.

\section{Summary and conclusions}

We explore the circumstellar {disc} of material {around} the AGB star L$_2$ Pup using radiative transfer models. In particular we aim to determine whether the observed sub-Keplerian rotation of the {disc} can be explained by radiation pressure acting in addition to centrifugal force and thermal pressure gradient to oppose gravity and whether this allows us to constrain the dust parameters. We draw the following main conclusions from this work. \\

\noindent 1. A thermal gas pressure gradient alone cannot explain the observed rotation profile without high temperatures that would be inconsistent with the state of the CO gas in the vicinity of L$_2$ Pup. \\

\noindent 2. Radiation pressure can drive sub-Keplerian rotation consistent with that observed. The dust population required to do this is mostly sensitive to grains in the range $0.1-1\,\mu$m, which dominate the opacity. There is hence a degeneracy between the dust-to-gas ratio and maximum grain size that implies a family of possible dust configurations that yield the observed rotation profile. Although this means that insights into the dust population from the rotation profile alone are limited, it can be coupled with other diagnostics. \\ 

\noindent 3. We run two classes of model, fixing either the maximum grain size or dust-to-gas mass ratio and allowing the other to vary radially. For models with fixed maximum grain size the dust-to-gas ratio has to increase radially (to increase the opacity, radiation pressure and hence deviation from Keplerian rotation). Similarly, for models with fixed dust-to-gas ratio the maximum grain size has to decrease radially.  \\

\noindent 4. The Stokes number of the grains that dominate the opacity is always much less than unity in our models, implying that the gas and dust are dynamically well coupled. We also validate this by solving for equilibrium solutions of the coupled dust-gas dynamics equations, finding tight coupling in the azimuthal velocity of gas and dust. These calculations also suggest that 0.$1\,\mu$m$-0.25$\,cm grains might be accelerated to high velocity ($\sim5$\,km/s). which would deplete the {disc} on ~20\,yr timescales. However the required dust replenishment rate of $\sim3\times10^{-9}$\,M$_{\odot}$\,yr$^{-1}$ is compatible, within uncertainties, with the observationally inferred mass loss rates for L$_2$ Pup of \cite{2002ApJ...569..964J} and \cite{2002MNRAS.337...79B}  \\

\noindent 5. Of our family of dust populations that yielded the correct rotation profiles, we computed SED models to compare with observations and further constrain the dust population. Our model with a fixed dust-to-gas ratio of $\delta=2.5\times10^{-3}$, maximum grain size of $240\,\mu$m at 6\,AU and radial variation of maximum grain size  according to equation \ref{fitequn}, is consistent with all observed data points longward of $10\,\mu$m (the regime dominated by dust emission). A second model with fixed $\delta=10^{-3}$ has the best fit beyond $10\,\mu$m according to a chi-square measure, although is not within uncertainties of the longest wavelength points. Generally though, it seems that lower than ISM dust-to-gas ratios give the best results. Note though that in reality all of the dust-to-gas ratio, maximum grain size and power law of the distribution can change radially, so there certainly are other good solutions in addition to those presented here.  \\

\noindent 6. Our models in this paper almost exclusively use \cite{2003ApJ...598.1017D} silicates. However we also tested iron rich and poor iron-magnesium silicates. We find that the higher absorption efficiency of iron-rich grains in the near-infrared \citep[e.g.][]{2006A&A...460L...9W} is compensated for by a higher scattering opacity of micron sized grains in that wavelength regime \citep{2008A&A...491L...1H}. The overall opacity therefore remains similar, but at larger radii in the {disc} the flux in the iron deficient case is actually slightly higher that the iron rich (since photons are scattered instead of absorbed). This means that for a grain population that only differs in its optical constants, iron deficient grains are more capable of driving sub-Keplerian rotation for lower dust masses, though the SED is not fit well by such a model. Iron rich grains give very similar results to \cite{2003ApJ...598.1017D} silicates. \\

Our results should motivate future studies that consider the dust-gas dynamics of the material {around} L$_2$ Puppis in a fully radiation hydrodynamic framework.  

\section*{Acknowledgements}
{We thank the anonymous referee for their review of the manuscript}. TJH is funded by an Imperial College Junior Research Fellowship. LD acknowledges support from the ERC consolidator grant 646758 AEROSOL. WH acknowledges support from the Fonds voor Wetenschappelijk Onderzoek Vlaanderen (FWO). This work was partly developed during and benefited from the MIAPP ``Protoplanetary discs and planet formation and evolution'' programme. This work used facilities provided by the DISCSIM project, grant agreement
341137 funded by the European Research Council under ERC-
2013-ADG.It also used the COSMOS
Shared Memory system at DAMTP, University of Cambridge operated
on behalf of the STFC DiRAC HPC Facility. This equipment is
funded by BIS National E-infrastructure capital grant ST/J005673/1
and STFC grants ST/H008586/1, ST/K00333X/1. DiRAC is part of
the National E-Infrastructure.  This paper makes use of the following ALMA data: ADS/JAO.ALMA\#2011.0.00277.S. ALMA is a partnership of ESO (representing its member states), NSF (USA) and NINS (Japan), together with NRC (Canada) and NSC and ASIAA (Taiwan) and KASI (Republic of Korea), in cooperation with the Republic of Chile. The Joint ALMA Observatory is operated by ESO, AUI/NRAO and NAOJ.

\bibliographystyle{mnras}
\bibliography{molecular}

%
%
%




\bsp	
\label{lastpage}
\end{document}